\documentclass[11pt]{article}
\usepackage{epsfig}
\usepackage{epstopdf}
\usepackage[usenames]{color}
\usepackage{graphicx}
\usepackage{amsmath}
\def\O{{\cal O}}
\def\la{\langle}\def\ra{\rangle}
\def\be{\begin{eqnarray}}\def\ee{\end{eqnarray}}
\def\lsim{\mathrel{\rlap{\lower3pt\hbox{\hskip1pt$\sim$}}
     \raise1pt\hbox{$<$}}} 
\def\gsim{\mathrel{\rlap{\lower3pt\hbox{\hskip1pt$\sim$}}
     \raise1pt\hbox{$>$}}} 
\def\le{ \begin{array}{ll}}\def\re{\end{array}}

\def\lear{ \left( \begin{array}{cc}}\def\rear{\end{array} \right)}

\def\le{ \left( \begin{array}{cc}}\def\re{\end{array} \right)}

\def\bi{\bibitem}

\def\Tr{\rm Tr}

\newcommand{\Rmnum}[1]{\expandafter\@slowromancap\romannumeral #1@}

\renewcommand{\thesection}{\arabic{section}}

\renewcommand{\theequation}{\arabic{equation}}

\def\lsim{\mathrel{\rlap{\lower3pt\hbox{\hskip1pt$\sim$}}
     \raise1pt\hbox{$<$}}} 
\def\gsim{\mathrel{\rlap{\lower3pt\hbox{\hskip1pt$\sim$}}
     \raise1pt\hbox{$>$}}} 

\def\la{\langle}
\def\ra{\rangle}
\def\bi{\bibitem}
\def\del{\partial}

\begin{document}

\centerline{\Large \bf Scale-Invariant Hidden Local Symmetry,}

\centerline{\Large \bf Topology Change and Dense Baryonic Matter }
\vskip 0.5cm
\begin{center}

{ Won-Gi Paeng$^a$\footnote{\sf e-mail: wgpaeng@ibs.re.kr},
Thomas T. S. Kuo$^b$\footnote{\sf e-mail: kuo@tonic.physics.sunysb.edu },
Hyun Kyu Lee$^c$\footnote{\sf e-mail: hyunkyu@hanyang.ac.kr} and
Mannque Rho$^d$\footnote{\sf e-mail: mannque.rho@cea.fr}}

\vskip 0.3cm

{\em $^a$Rare Isotope Science Project, Institute for Basic Science, Daejeon 305-811, Korea}

{\em $^b$Department of Physics and Astronomy, Stony Brook University, Stony Brook, New York 11794, USA }

{\em $^c$Department of Physics, Hanyang University, Seoul 133-791, Korea }

{\em $^d$Institut de Physique Th\'eorique, CEA Saclay, 91191 Gif-sur-Yvette c\'edex, France }

\vskip 0.3cm
{\ (\today)}
\vskip 1.2cm
\end{center}
\centerline{\large\bf ABSTRACT}
When scale symmetry is implemented into hidden local symmetry in low-energy strong interactions to arrive at a scale-invariant hidden local symmetric (HLS) theory, the scalar $f_0(500)$ may be interpreted as pseudo-Nambu-Goldstone (pNG) boson, i.e., dilaton, of spontaneously broken scale invariance,  joining the pseudo-scalar pNG bosons $\pi$ and the matter fields $V=(\rho,\omega)$ as relevant degrees of freedom. Implementing the skyrmion-half-skyrmion transition predicted at large $N_c$ in QCD at a density roughly twice the nuclear matter density found in the crystal simulation of dense skyrmion matter, we determine the intrinsically density-dependent (IDD) ``bare parameters" of the scale-invariant HLS Lagrangian matched to QCD at a matching scale $\Lambda_M$. The resulting effective Lagrangian, with the parameters scaling with the density of the system, is applied to nuclear matter and dense baryonic matter relevant to massive compact stars by means of the double-decimation renormalization-group  $V_{lowk}$ formalism. We satisfactorily post-dict the properties of normal nuclear matter and more significantly {\it predict} the EoS of dense compact-star matter that quantitatively accounts for the presently available data coming from both the terrestrial and space laboratories. We interpret the resulting structure of compact-star matter as revealing how the combination of hidden-scale symmetry and hidden local symmetry manifests itself in compressed baryonic matter.

\newpage

\section{Introduction}
In a preceding note~\cite{LPR15}, the notion that  the $f_0(500)$, the lowest scalar listed in the particle data booklet, is a dilaton arising from the spontaneous breaking of scale invariance in QCD~\cite{CT} was implemented into hidden local symmetry (HLS)~\cite{HY:PR} of the light-quark vector mesons $V_\mu= (\rho_\mu, \omega_\mu)$ that embodies the non-linear realization of chiral symmetry into a scale-invariant hidden local symmetry theory ($s$HLS for short) and the resulting Lagrangian was subjected to the vacuum change due to the density of baryonic matter.

In this paper, we confront the resulting formalism with what's known of normal nuclear matter and make predictions on properties of dense matter appropriate for massive compact stars.

 Since the basic premise for the effective Lagrangian that we shall employ, $s$HLS, is fully expounded in \cite{LPR15}, we shall eschew details and limit ourselves here only to what are essential for the calculations that we make. We shall follow closely the procedures given in \cite{LPR15}. The only issue that was not given an adequate comment in \cite{LPR15} is the place of an infrared (IR) fixed point postulated in \cite{CT} in QCD with the number of flavors $N_f\sim 3$ as needed in nuclear phenomena. It is perhaps worth making a brief remark on it.
  It is argued in \cite{CT} that the notion that QCD has an IR fixed point for $N_f=3$ with the resulting ``scale-chiral symmetry" solves some of the long-standing puzzles in particle physics that involve ``light" scalar excitations. For instance, phrased in terms of a scale-chiral counting rule generalizing the chiral counting rule of chiral perturbation theory, it gives a surprisingly  simple explanation of  the $\Delta I=1/2$ rule, accounts for the mass and width of the scalar $f_0(500)$ etc. As stressed in \cite{LPR15}, it could also resolve long-standing conundrums in nuclear physics involving a low-mass scalar. Unfortunately, however, there is, so far, no convincing proof that the three-flavor QCD has an IR fixed point: Neither lattice nor model-independent approaches have uncovered it. This is in contrast to QCD at $N_f\sim 8$ being studied for dilatonic Higgs model for going beyond the Standard Model (for a recent summary, see \cite{yamawaki-dilaton})\footnote{Private communication from Koichi Yamawaki}.  This does not imply that an IR fixed point advocated in \cite{CT} is ruled out. As argued in \cite{CT}\footnote{Also in private communication from R. Crewther and L. Tunstall}, an IR fixed point at which scale-chiral symmetry is realized in Nambu-Goldstone mode has not yet been probed by the lattice work.

 In this paper, we take the point of view that in dense matter, the scale-chiral symmetry of the sort advocated by \cite{CT} could be present as an ``emergent symmetry."  This is in some sense similar to hidden local symmetry which plays an equally important role in our calculation.  The notion of hidden local symmetry which gives the famous ``VD(vector dominance)" and KSRF relation makes sense only if the vector meson $V_\mu$ is light. There are two known cases where the ``lightness" of $V_\mu$ is realized. One is the presence of the vector manifestation (VM) fixed point at which the vector meson mass goes to zero as does the pion mass (in the chiral limit)~\cite{HY:PR}. The other is supersymmertric QCD in certain parameter space~\cite{sqcd}. In what follows, the VM fixed point, emerging at high density, will play a key role. In a similar vein, a scalar of $\sim 600$ MeV fluctuating around an IR fixed point will figure crucially in the equation of state (EoS) for dense matter.

 It is rather intriguing that the two symmetries we are combining, i.e., scalar symmetry and (vector) local symmetry , are hidden in baryonic matter (as in the beyond-the-Standard-Model regime~\cite{yamawaki-dilaton}) and seem to emerge at high density.

\section{Scale-invariant HLS Lagrangian}
The Lagrangian we will consider, $s$HLS, simplified from \cite{LPR15}, takes the form
\be
 {\cal L}_{sHLS} (U,\chi,V_\mu) & \approx&
  {\cal L}_{HLS}^{(2)} (\frac{\chi}{f_{0\sigma}})^2 +  \frac{f_{0\pi}^2}{4} (\frac{\chi}{f_{0\sigma}})^{3} {\Tr}(MU^\dagger +h.c.) + \cdots \nonumber\\
   &+& \frac 12 \del_\mu\chi\del^\mu\chi + V(\chi) \,\label{LAG}
 \ee
 where $\chi=f_\sigma e^{\sigma/f_\sigma}$ is the ``conformal compensator" field with $\sigma$ the non-linear dilaton field and $f_\sigma=\la\chi\ra$ the vev (either matter-free or in-medium),  the chiral field $U$ consists of (L,R) fields as $U=e^{i\frac{2 \pi}{f_\pi}}=\xi^\dagger_L\xi_R$ with $\pi=\frac 12 \vec{\tau}\cdot\vec{\pi}$,
 $M$ is the quark-mass matrix representing chiral symmetry breaking which also breaks scale symmetry,  $f_{0\sigma}$ is the medium-free-vacuum expectation value $\la 0|\chi|0\ra$ and $V(\chi)$ is the dilaton potential that encodes the spontaneous and explicit breaking of scale invariance.  For simplicity, the HLS Lagrangian is given to ${\O}(p^2)$,  with the ellipsis standing for higher scale-chiral order terms. Note that we are taking the approximation $c_i\approx 1$ in the notation of Ref. \cite{CT}.  The potential $V(\chi)$ contains several unknown constants in \cite{CT} of which we do not need their specific forms for our analysis.
 \subsection{Intrinsic density dependence (IDD) of ``bare" parameters of $s$HLS}
 In order to confront the Lagrangian (\ref{LAG}) with nuclear matter and high density matter, there are three indispensable ingredients to consider. First, the baryon degrees of freedom have to be incorporated. Second, the ``bare" parameters of the effective Lagrangian need to be matched to QCD. Third, strong correlations between nucleons, including possible phase changes,  have to be included as one goes up in density.

 All three could in principle be handled -- at least in some approximations such as large $N_c$ -- using skyrmion description of baryons and baryonic matter~\cite{LPR15}. Some progress has been made in this direction~\cite{maetal} but the mathematics required is still too daunting to arrive a reliable result. We shall therefore put baryons explicitly ``by hand" in scale-chiral symmetric way. Let us call the baryon-field-implemented Lagrangian $bs$HLS for short. Since high density, $n\sim (5-7)n_0$ (where $n_0$ is the nuclear matter density), going toward chiral transition is involved, the ``bare" parameters need to have contact with QCD parameters. This will be done by matching the correlators of the $bs$HLS Lagrangian to those of QCD at an appropriate matching point $\Lambda_M$ lying below the chiral scale $\Lambda_\chi\sim 1$ GeV, say, at about the $\rho$ mass. It is important to note that the matching endows the ``bare" parameters of EFT Lagrangian with dependence on the quark condensate $\la\bar{q}q\ra$, the gluon condensate $\la G^2\ra$ etc. Since those condensates depend on the ``vacuum," they will of course depend on density which modifies the vacuum if the EFT Lagrangian is embedded into a medium. The crucially important point in our development is that {\it the density dependence involved here is {\it intrinsic} of QCD, to be distinguished from the density dependence coming from (mundane) nuclear many-body correlations.} This density dependence -- that will play a key role in what follows -- will be referred, as in \cite{LPR15}, to as ``intrinsic density dependence" (IDD for short).
 \subsection{Double-decimation RG procedure}
 Now given the EFT Lagrangian endowed with the IDDs, nuclear dynamics is treated by renormalization-group decimation from the matching scale $\Lambda_M$ down to the appropriate low-energy scale where the processes we are interested in take place. For this, a highly versatile tool is  the $V_{lowk}$ strategy~\cite{bogner03,bogner03b}. We will exploit it in this paper.  A convenient -- and successful -- procedure in nuclear physics is the ``double decimation" RG flow described in \cite{BR:DD}. In fact, this procedure was used for the first time for the EoS for massive stars in \cite{dongetal}. In this paper, we will improve on it both in concept and in numerics, assuring consistency with the scale-chiral symmetry adopted in \cite{LPR15}.

 In the $V_{lowk}$ framework, the double decimation consists of the first step from $\Lambda_M$ to the scale at which $V_{lowk}$ is obtained. The second step is to decimate to the Fermi sea around which fluctuations are computed to take into account multi-body correlations. This is equivalent to fluctuating around the Landau Fermi-liquid fixed point~\cite{kuo-brown-fermiliquid}. It is in doing these decimation calculations using $V_{lowk}$ that information from a topology change encoded in the skyrmion crystal treatment of dense matter enters. This transition involves no local order-parameter field and hence may not belong to the Ginzburg-Landau-Wilson paradigm but as will be seen, has a drastic impact on the EoS in compact-star matter. While the topology change that takes place in the skyrmion crystal is, strictly speaking, valid only in the large $N_c$ limit of QCD, it seems quite universal, visible already in the structure of the alpha particle with four nucleons~\cite{multifacet,manton-sutcliffe}.  It is thus highly plausible that such a  half-skyrmion topological structure could be present in dense matter, say, above nuclear matter density. What is done in this paper is that this feature of changeover from skyrmions to half-skyrmions in the soliton description  is translated into the bare parameters of the effective Lagrangian, in terms of changes in IDDs.  It effectively demarcates the EFT Lagrangian into two density regimes, one for (I) $n\lsim n_{1/2}$ and the other for (II) $n\gsim n_{1/2}$. The former (I) entails the ``bare" parameters of the Lagrangian that carry the density dependence referred to as ``IDD$_{pNG}$" where the (pseudo-)NG bosons figure and the latter (II)  ``IDD$_{matter}$" in which the matter fields $V_\mu$ intervene. An interesting observation made in Higgs physics~\cite{yamawaki-dilaton} where also both hidden scale symmetry and hidden local symmetry enter is that the properties of (techni)vector mesons are scale-invariant. Intriguingly, it turns out also in dense matter that the $\rho$ meson properties are scale-invariant, controlled  by the VM fixed-point structure.

\section{Symmetry Energy, Tensor Forces and Topology}
One of the most interesting observables in dense matter is the ``symmetry energy factor" $S$ defined in the energy $E$ per particle of nucleus consisting of $P$ protons and $N$ neutrons, i.e., $A=P+N$,
\be
E(n,x)=E_0 (n,0)+S(n) x^2 +\cdots,\label{E}
\ee
with $x=(N-P)/A$. Here $n$ stands for baryon number density and the ellipsis stands for higher-power terms in $x$. The quadratic approximation is known to be reliable, so we focus on $S$. As is well-known, the symmetry energy is the quantity, representing neutron excess of the system, that plays a key role in the EoS of compact stars. It is this quantity that is strongly influenced by topology in the skyrmion picture, manifested through the nuclear tensor forces.

As mentioned above, a robust feature of skyrmion description of nucleonic matter -- that we shall exploit in what follows --  is that there is a ``changeover" from a state of skyrmions to a state of half-skrymions at some matter density denoted $n_{1/2}$. Its presence in the skyrmion framework is remarkably independent of the degrees of freedom involved and is quite insensitive to the parameters of the Lagrangian. It is present in the Skyrme model with pions only as well as in $s$HLS models with the vectors and/or the dilaton~\cite{multifacet}. Precisely at which density the changeover takes place is, however, model-dependent and cannot be pinned down precisely in the present state of formulation. However the density at which the half-skyrmion appears, $n_{1/2}$, is found to be insensitive to the dilaton mass, the most uncertain quantity in the calculation. This feature is seen in the model in which skyrmions are put on crystal lattice~\cite{maetal,LPRV-pionvelocity}. In what follows, {\it our basic premise will be that in terms of the skyrmion picture justified at high density and for large $N_c$, half-skyrmions could appear at some density in the vicinity of $\sim 2n_0$.} What ensues is a striking consequence on the symmetry energy

Since the phenomenon considered is quite generic, more or less independent of the degrees of freedom involved,  we can address the matter using the simplest model, i.e., the Skyrme model~\cite{skyrme} that consists of two terms, the current algebra term and the Skyrme quartic term implemented with the conformal compensator field. It corresponds to dropping the vector meson fields and putting the Skyrme quartic term -- which is of scale dimension 4 and hence scale-invariant -- in place of the ellipsis  in Eq.~(\ref{LAG}). We expect the result to be qualitatively the same with the more realistic Lagrangian (\ref{LAG}).

With skyrmions put on crystal, the easiest way to compute the symmetry energy is to rotationally quantize A-neutron skyrmion matter, which corresponds to calculating $S$ from (\ref{E}) for $x=1$~\cite{LPR-SE}.  It is given by
\be
S\approx \frac{1}{8\lambda_I}.\label{lambda_i}
\ee
Here $\lambda_I$ is the isospin moment of inertia of ${\cal O} (N_c)$ given by the space integral over the single cell of the hedgehog configuration $U_0$ and the dilaton configuration. In the presence of vector mesons,  the integral will also involve the mesons's  classical configurations.  It is of the leading order in $N_c$, with fluctuation corrections suppressed by $1/N_c$.

The striking feature of the symmetry energy factor (\ref{lambda_i}) turns out to  be a cusp structure at the changeover density $n_{1/2}$ - which comes out at $n_{1/2}\sim (1.3 -2.0)n_0$. The numerical calculation of Equation (\ref{lambda_i}) reveals that the $S$ decreases monotonically as density increases toward $n_{1/2}$ and then turns up and monotonically increases after $n_{1/2}$.

Now it may be that the method anchored on crystal is not applicable to low-density matter. Furthermore nuclear matter at equilibrium density is known to be in Fermi liquid. Therefore  one might object to applying the crystal skyrmion description not too far above the nuclear matter density. However it turns out that the cusp structure at a density at $\sim 2n_0$ is not an artifact of crystal background and can be trusted.  In fact what is highly nontrivial is that this feature can be easily reproduced by the microscopic structure of the tensor forces, in particular, the effect on the tensor forces of the topological change at $n_{1/2}$. For this, we use the fact that the symmetry energy is dominated by the tensor forces~\cite{tensor forces}. We first write the effective Lagrangian (\ref{LAG}) that implements the topology change at $n_{1/2}$ in the skyrmion description. To do this, we divide the density regime into two regions --R(egion) I and II -- with the demarcation at $n_{1/2}$,
\be
&&{\rm R(egion)-I}:  0<n< n_{1/2},\label{regionI}\\
&&{\rm R(egion)-II}: n_{1/2}\leq n\leq n_c.\label{regionII}
\ee
As described in \cite{LPR15}, we can translate the topology change into scaling (that is, IDD) properties of the parameters of the Lagrangian (\ref{LAG}) in the two regions. The principal parameters involved are the decay constants $f_{\pi,\sigma}$ and  masses $m_{\pi,\sigma}$ of the pseudo-Nambu-Goldstone bosons, the coupling constants $g_{\rho,\omega}$ and masses $m_{\rho,\omega}$ of the hidden gauge fields etc. The specific parametrization that we extract from the strategy detailed in \cite{LPR15} and will be used for the $V_{lowk}$ approach presented in Section \ref{vlowk} is described in the next section. Here we make use of it in showing how the cusp in $S$ can be understood in the given framework. In this approach, one first constructs nuclear potentials in terms of the exchange of the meson degrees of freedom given in the Lagrangian with the scaling parameters. Apart from the IDDs in the Lagrangian, this is essentially what is done in nuclear chiral perturbation theory. Now in terms of our $s$HLS Lagrangian, the tensor forces $V_T$ consist of $\pi$ and $\rho$ exchanges, $V_T=V_T^\pi + V_T^\rho$. The notable feature of $V_T^{\pi,\rho}$ is that the two contributions, having the same radial form with different masses, come with the opposite sign. Thus the net tensor force involves a crucially important cancelation between the two components, which depends on the scaling properties of the two components. As will be seen in the next section, the details are a bit involved, but the qualitative feature is simple.

As noted, the prominent feature of the net tensor force at $n_{1/2}$ is the abrupt change in the slope. In R-I, the pion tensor is  almost completely unaffected by density within the range of density we are considering,  due to what one might  interpret as the protection by chiral symmetry. This has been numerically confirmed up to $\simeq 5n_0$. See Appendix \ref{appen_tensor}.  The $\rho$ tensor, on the other hand, gets enhanced as density increases due to the dropping of its mass. Since it comes with the sign opposite to the pion tensor, it cancels part of the pionic tensor. The net effect is then the tensor force becomes weaker as density increases. This tendency is in agreement with a variety of observations in nuclei, most spectacular of which is the long life-time for carbon-14~\cite{c14}, i.e., the C-14 dating.  (It should be mentioned that short-range three-body forces, present as contact interaction in chiral perturbation theory, could do the same suppression of the Gamow-Teller matrix element involved. As explained in \cite{holt-rho-weise}, however,  this does {\it NOT} represent a different mechanism to that of \cite{c14}. It may be said that most, if not all, of the effect of the contact 3-body forces,largely responsible for the suppression of the Gamow-Teller matrix element, is encoded in the IDD included in \cite{c14}.)  The weakening of the net tensor force continues up to the changeover density $n_{1/2}$. At $n_{1/2}$,  the tensor force stops decreasing, turns over and starts increasing, with the pion tensor becoming dominant. There are two mechanisms at work here. One is that the change of parameters that takes place at $n_{1/2}$ strongly suppresses the overall strength of the $\rho$ tensor force although the mass continues dropping. The other is that in R-II, the candidate order parameter for chiral symmetry is a four-quark condensate with the bilinear quark condensate suppressed (it goes to zero at $n_{1/2}$ in the skyrmion crystal). And the four-quark condensate is found to be strongly suppressed in Region II~\cite{maetal}. Interpreted in terms of a GMOR relation for in-medium pion, this would imply, since the in-medium decay constant remains more or less un-scaling in density in R-II, that the pion mass must then decrease. As a consequence, the pion tensor must become stronger, even further enhanced over and above the free-space value. This will facilitate pion condensation, as expected at high density in crystal form. Given the abrupt change in the tensor force at $n_{1/2}$, the cusp structure in the symmetry energy found in the skymion model follows in an immediate way as explained below.

In summary, there is a change in the slope of the symmetry energy factor $S$ at $n_{1/2}$,  a semi-classical result (in the sense of large $N_c$ effect) which is a robust feature in the framework of $s$HLS theory. How it manifests in nature requires a sophisticated treatment of many-body theory. What follows in this paper is a detailed analysis of this feature in the RG-implemented $V_{lowk}$ approach which takes into account high-order correlations encoded in Landau-Fermi liquid theory.

\section{Topological Demarcation of Density Regimes}
\subsection{Intrinsic Density Dependence (IDD)}\label{scaling}
In this section we specify the effective $s$HLS Lagrangian that is endowed with the density dependence IDD inherited from QCD at the matching scale $\Lambda_M$.    As announced, we deal with two density regimes -- Region I and II -- when the system is embedded in medium. That there can be two regimes  demarcated at a density above $n_0$ is neither indicated by a general QCD argument nor by model-independent effective field theory arguments. It is however predicted in the skyrmion description of dense matter which is strictly valid in the large $N_c$ limit and at high density. We take this into account by interpreting, as described above, the demarcation as the changes in the density dependence of the effective Lagrangian that is applicable to the $V_{lowk}$ approach.

The Lagrangians (\ref{L_I}) and (\ref{L_II}) applicable in R-I and R-II, respectively, are written in Lorentz -invariant form.  One may object to their form saying that they should actually take $O(3)$ covariant form in medium since the Lorentz symmetry is spontaneously broken.  In fact the $O(3)$ covariant HLS Lagrangian was written down before~\cite{lorentz}. However in the scheme we are using with the correlators, the density dependence of the bare parameters of the Lagrangian is in ``vacuum-specific condensates."  These do not intervene in spontaneous breaking of Lorentz symmetry~\cite{kaempfer}. The symmetry breaking that breaks the $O(4)$ symmetry comes in RG decimations \`a la $V_{lowk}$ with the given Lagrangians.  Furthermore  in the hidden-scale-HLS framework, as density goes above $n_{1/2}$, the quark condensate, while supporting chiral density wave, goes to zero on average and the vector mass drops rapidly toward the VM fixed point.  When these two phenomena take place, the Lorentz symmetry breaking decreases surprisingly rapidly~\cite{lorentz}. Thus the pion velocity, for instance,  approaches 1 quickly. We should point out that the situation is totally different in the absence of the vector meson with the VM fixed point~\cite{son-stephanov}. The relativistic mean field approach with the Lagrangian with density-dependent parameters, popularly used in nuclear theory circles, is justified along this line of reasoning at high density.

\subsubsection{Region-I}
Consider first Region-I (\ref{regionI}). This is the normal nuclear matter phase,  extrapolated to density $n_{1/2}$ which can be described in a multitude of phenomenologically reliable models. Currently most popular is the chiral perturbative approach, i.e., two-flavor $\chi$PT$_2$. The $V_{lowk}$ approach can be considered as an improved version of  $\chi$PT$_2$, in that one universal IDD intervenes and improves on the phenomenology in the vicinity of nuclear matter where data are available.

To be specific while preserving simplicity, we write the in-medium  ``bare" $bs$HLS Lagrangian in a linearized form
    \be
{\cal L}_I &=& \overline{N}[i\gamma_\mu(\del^\mu+igV^\mu)-m_N^*+g_{\sigma}\sigma]N
 - \frac14 V_{\mu\nu}^2+\frac{{m^*_V}^2}{2}V^2 \nonumber\\
 &+& \frac 12(\del_\mu\sigma)^2 -\frac{{m_\sigma^*}^2}{2}\sigma^2 +\frac 12 \del^\mu\vec{\pi}\cdot\del_\mu\vec{\pi} -\frac 12 {m_\pi^*}^2 {\vec{\pi}}^2  +{\cal L}_{\pi m} + \cdots\label{L_I}
 \ee
 where the ellipsis stands for possible terms that are of higher order in chiral-scale counting and of higher fields and $V_\mu=\vec{\tau}\cdot\vec{\rho}_\mu +\omega_\mu$ assumed to be flavor-$U(2)$ symmetric.  Since the flavor $U(2)$ symmetry for the vectors $V_\mu$ seems to be fairly good in the matter-free vacuum, it should hold also in low-density regime, i.e., R-I.  (At high density in R-II, however, we will find that the $U(2)$ symmetry must break down~\cite{PLRS-nucleonmass}.)  ${\cal L}_{\pi m}$ stands for the pion-matter and pion-$\sigma$ couplings. The matching of the EFT Lagrangian to QCD renders the pion decay constant $f_\pi$ and the dilaton decay constant $f_\sigma$ dependent on the QCD condensates ${\cal C}$, i.e.,  $\la\bar{q}q\ra$, $\la G^2\ra$ etc. Since the condensates reflect the vacuum structure, in medium, the decay constants depend on density, which will be denoted with an asterisk, $f_{\pi,\sigma}^*$. As stated,  {\it this density dependence is an intrinsic property of the QCD vacuum structure, to be distinguished from density dependence that is due to standard nuclear many-body correlations.} This distinction arises from the strategy of matching EFT to QCD.

 Following the reasoning given in \cite{LPR15}, we can relate the in-medium decay constants as
 \be
 f_\pi^*/f_{0\pi}\approx f_\sigma^*/f_{0\sigma} \equiv \Phi_I (n)\label{phi_i}
 \ee
 where $f_{0\pi,\sigma}$ are the decay constants in the matter-free vacuum.\footnote{ Given the ``$c\approx 1$ approximation" made in locking chiral symmetry to scale symmetry~\cite{LPR15}, we use approximate equality instead of equality.} This follows from the nature of Nambu-Goldstone bosons reflecting the locking of the chiral symmetry to the scale symmetry, that is, IDD$_{pNG}$. The pion and dilaton decay constants depend on QCD condensates ${\cal C}$, the former on the quark condensate \`a la GMOR and the latter on both the quark condensate and gluon condensate~\cite{LPR15,CT}. Since there is no lattice calculation in dense medium, the scaling function $\Phi$ is not really known from QCD proper. For low density, one may resort to chiral perturbation theory $\chi$PT$_2$. More pertinently -- and fortunately -- there is information from experiments where the pion decay constant is measured up to $n_0$, e.g., in deeply bound pionic system. One immediate consequence of IDD$_{pNG}$ is the d(ensity)-scaling of the pion mass
 \be
 m_\pi^*/m_{\pi}\approx \Phi_I^{1/2}.\label{pi-mass}
 \ee
 How the dilaton mass d-scales is more complicated. We will return to it later.

 As for the properties of, and coupling, to the matter fields, one needs to consider the IDD$_{matter}$, that is, due to the matching of the vector and axial vector correlators. However as argued in \cite{LPR15} and elsewhere based on phenomenology, to the order we are considering, the IDD$_{matter}$ can be ignored in R-I,  so we can focus only on IDD$_{pNG}$ effects.\footnote{The matching of the vector and axial vector correlators does make the pion decay constant $f_\pi$ inherit the quark and gluon condensates from QCD but their effects are negligible. It cannot account for vanishing pion decay constant as the quark condensate is dialled to zero. It requires a subtle role of quadratic divergence in the pion loops in RG decimation. Furthermore $f_\sigma$ -- that locks scale symmetry to chiral symmetry -- cannot enter into the correlators of the isovector currents we have for IDD$_{matter}$. On the contrary, we will see later the situation is entirely different in Region-II. } This yields
 \be
 m_N^*/m_N\approx m_V^*/m_V\equiv \Phi_I.\label{matter-mass}
 \ee
 To the same approximation, the hidden gauge coupling $g$ and  the $\sigma NN$ coupling $g_\sigma$ do not d-scale
 \be
 g^*/g\approx g_\sigma^*/g_\sigma\approx 1.\label{coupling}
 \ee
 On the contrary, the pion-NN coupling in $g_{\pi NN}(\bar{N}\frac 12 \vec{\tau}\cdot\vec{\pi}\gamma_5) N$ d-scales\footnote{It should be noted that the conformal compensator trick used in this paper works differently between the linear $\pi$-nucleon coupling which is used for (\ref{gpinn}) and the nonlinear coupling that figures in the $bs$HLS Lagrangian. In the latter, the axial coupling constant $g_A$ will not d-scale and hence neither will $g_{\pi NN}$. This has to do with a well-known problem of the so-called ``quenching of $g_A$ in nuclei" that comes from the role of the $\Delta$ resonance in the baryon sector. The $g_A$ obtained in the linear coupling accounts for the role of the $\Delta$ that is integrated out from the $bs$HLS Lagrangian. This is an old story that dates back to 1974 with the quenching of $g_A$ by a $\Delta$-hole mechanism. See, e.g., \cite{MR74}. In this paper we will use this scaling which is not properly included in the IDD$_{pNG}$ but is required for consistency.}
 \be
 g_{\pi NN}^*/g_{\pi NN} \approx \Phi_I.\label{gpinn}
 \ee
 This implies, by the low-energy theorem known as Goldberger-Treiman relation,
 \be
 g_A^*/g_A\approx \Phi_I.
 \ee
 Finally we turn to the dilaton mass $m_\sigma^*$. As discussed at length in \cite{LPR15}, the dialton being a pseudo-Goldstone scalar with explicit scale-symmetry breaking due to an intricate interplay, un-understood yet, of the trace anomaly and the current quark mass, we are unable to determine with confidence the d-scaling of the dilaton mass with the dilaton potential of \cite{CT}. If however one took the dilaton potential of the Coleman-Weinberg-type log potential just to have an idea, one would obtain
 \be
 m_\sigma^*/m_\sigma\approx \Phi_I.\label{sigma-mass}
 \ee
 In R-I, a reasonable parametrization that we shall use is
 \be
 \Phi_I (n)\approx \frac{1}{1+c_I\, n/n_0}.\label{PhiI}
 \ee
 The value of $c_I>0$ used in numerical analysis will be given in Section \ref{eos}.

 Though not highly rigorous, this is supported up to nuclear matter density~\cite{BR-walecka-kcon} by IDD-implemented Walecka-type mean-field, so we will assume it in the numerical analysis given below. This completely determines the bare Lagrangian (\ref{L_I}). Only one d-scaling function $\Phi_I$ is to be determined and this can be done by resorting to pionic nuclear systems and/or chiral perturbation theory. For quantitatively accurate agreement with Nature, however, a small fine-tuning on $c_I$ will be required in Section \ref{eos}.

 \subsubsection{Region-II}\label{region-II}
 In this region, there is no guidance either from experimental data or from trustful theory -- except for the hidden local symmetry prediction given below in (\ref{VMHLS}). This makes a precise determination of the effective Lagrangian problematic. Thus our approach is highly exploratory and uncertain. What is clear is that the density dependence of the parameters must undergo drastic modifications as the system goes across the changeover point $n_{1/2}$: First chiral perturbation theory, formulated to work well up to nuclear matter density, most likely breaks down at some high density in Region-II. This is because chiral perturbation theory makes sense in small-$k_F$ expansion whereas Fermi liquid fixed point approach relies on small $1/k_F$ expansion~\cite{shankar}. Second the local $U(2)$ symmetry assumed in R-I  is likely to break down. Third, most significantly,  in hidden local symmetry for the $\rho$ meson which would be more justified as the vector-meson mass drops to the level of pNG bosons, there is the vector manifestation (VM) of hidden local symmetry, at the approach to which the mass d-scale to zero as
 \be
 m_\rho\sim g_\rho\sim \la\bar{q}q\ra\rightarrow 0\label{VMHLS}
 \ee
as $\la\bar{q}q\ra\rightarrow 0${, where we define $g_\rho$ as the hidden local gauge coupling for $\rho$ to distinguish it from $g_\omega$ for $\omega$.} What is significant in this behavior is that it is the hidden gauge coupling $g_\rho$ -- which is un-scaling in R-I unaffected by IDD$_{pNG}$ -- that plays an important role. Similarly the pion decay constant vanishing only very near the density at which chiral symmetry is restored, hence in R-II approaching the density that drives the system to the VM fixed point, is intricately connected to the matching process~\cite{HY:PR}.
  This means that IDD$_{matter}$ must become operative in R-II; (\ref{LAG}) from a phenomenological point of view, were the parameters of Region I to continue to higher density much above $n_0$, then the symmetry energy factor would become ``supersoft'' at a density $n\gsim (3-4)n_0$ which would require modification to gravity theory~\cite{bao-an}.

There is also a possibility that the Fermi-liquid structure, assumed to hold in R-I, breaks down in R-II. This possibility will not be considered in this paper.

To account for a rapid changeover at $n_{1/2}$ in $s$HLS, we take the d-scaling for the $\rho$ vector meson in R-II to be consistent with the VM
\be
m^*_\rho/m_\rho \,{ \propto }\, g_\rho^*/g_\rho \equiv \Phi^\rho_{II}\, .\label{vmass-II}
\ee
Approaching the VM fixed point, we take the linear density scaling
\be
  \Phi^\rho_{II} (n) \approx (1-c^\rho_{II}n/n_0)\label{PhiII}
\ee
 with $c^\rho_{II}$ will be fixed to give the chiral restoration density, for rough estimate,  $n_c \sim (6-7)n_0$.

The density $n_{1/2}\sim 2n_0$ may be a bit too far from the VM fixed point for this d-scaling (\ref{vmass-II}) to be quantitatively accurate, but one can take this as expanding around the VM fixed point  as was done for kaon condensation that takes place at $n\sim 3n_0$~\cite{BLPR-kcon}. An approximately same critical density -- near $n_{1/2}$ -- is arrived at by expanding around equilibrium nuclear matter treated as the Fermi-liquid fixed point~\cite{BR-walecka-kcon}.

{If the local $U(2)$ symmetry for $(\rho, \omega)$ were good in R-II as it seems to be in R-I, one could use the same reasoning given above for $\rho$. However there is nothing to indicate that the symmetry would hold there. For instance, the reasoning that goes into the VM fixed point for the $\rho$ based on correlators as given in \cite{HY:PR} does not apply to the $\omega$ meson. In fact if one assumes $U(2)$ symmetry and let the $\omega$ behave in the same way as the $\rho$ in R-II with the VM property (\ref{PhiII}), both symmetric nuclear matter and neutron matter become unstable just above the demarcation density $n_{1/2}$.  This feature, shown in Appendix \ref{AR-II}, is the first clear message {\it within the framework developed in this paper} that $U(2)$ symmetry could be badly broken at high density.  We shall therefore relinquish the $U(2)$ hidden local symmetry for the vector mesons and treat the $\rho$ in $SU(2)$ HLS  and the $\omega$ in $U(1)$ HLS as in \cite{PLRS-nucleonmass}.

The $\omega$ mass formula takes the same Higgsed mass as that of  $\rho$,
\be
m^2_\omega=f^2_\omega g^2_\omega\,,\label{omegamass}
\ee
where {$m_\rho^2 = f_\rho^2 g_\rho^2$ and $f_\omega$ is the $U(1)$ analog to $f_\rho=\sqrt{a_\rho} f_\pi$}\footnote{Phenomenologically, $a_\rho$ is determined to be $\sim 2.1$ in the matter-free space\cite{HY:PR}.}. In analogy to the case of $\rho$, we define $a_\omega$ as
\be
f_\omega=\sqrt{a_\omega} f_\pi.
\ee
Now we do not know how $m_\omega$ scales apart from the IDD$_{pNG}$ factor $f_\sigma^\ast$.  In fact, neither $f_\omega$ nor $g_\omega$ is known in medium~\footnote{Note in the vacuum the near mass degeneracy of $\rho$ and $\omega$ gives the hint that $g_\rho\approx g_\omega$ and $a_\rho\approx a_\omega$.}. In what follows in confronting Nature, we will rely on Nature to guide us in arriving at the properties of $\omega$ at high density.}

For the d-scaling of other quantities, we again resort to qualitative features found in the skyrmion crystal simulation focusing on the skyrmion-half-skyrmion changeover~\cite{maetal}. They are
\begin{itemize}
\item In-medium nucleon mass $m_N^*$ goes like $f_\pi^*$ which is consistent with the large $N_c$ property $m_N^*\sim ef_\pi^*$ where $e\sim {\cal O} (N_c^{1/2})$ is related to the scale-invariant Skyrme term in the Skyrme Lagrangian, hence non-d-scaling. Somewhat surprisingly, the pion decay constant remains roughly non-d-scaling after $n_{1/2}$ until very near the chiral restoration point. Therefore we think it reasonable to take
    \be
    m_N^*/m_N\approx m_\sigma^*/m_\sigma\approx f_\sigma^*/f_{0\sigma}\approx f_\pi^*/f_{0\pi}\approx \kappa\label{m-II}
     \ee
    where $\kappa\leq 1$ is more or less non-d-scaling constant up to near the chiral restoration density at which it could drop to zero.
\item If one assumes that the Goldberger-Treiman-like relation with the dilaton holds, i.e., $m_N^*\approx g_\sigma^* f_\sigma^*$ ~\cite{LPR15}, then it is a good approximation to take
\be
g_\sigma^*/g_\sigma\approx  {\rm constant}\approx 1.
\ee
This completes the density dependence of the Lagrangian in R-II,
\be
{\cal L}_{II} &=& \overline{N}[i\gamma_\mu(\del^\mu+ig^*V^\mu)-m_N^*+g_{\sigma}\sigma]N
 - \frac14 V_{\mu\nu}^2+\frac{{m^*_V}^2}{2}V^2 \nonumber\\
 &+& \frac 12(\del_\mu\sigma)^2 -\frac{{m_\sigma^*}^2}{2}\sigma^2 +\frac 12 \del^\mu\vec{\pi}\cdot\del_\mu\vec{\pi} -\frac 12 {m_\pi^*}^2 {\vec{\pi}}^2  +{\cal L}_{\pi m} + \cdots\label{L_II}
 \ee

\end{itemize}
If $U(2)$ symmetry held for the vector mesons, there would be only two parameters in Region II, the constant $\kappa$ and the d-scaling factor $\Phi^\rho_{II}$ that should go to zero as the VM fixed point is approached. Recall that in Region I, there is only one d-scaling function $\Phi_I$.
As shown in Appendix \ref{AR-II}, the symmetry is broken in medium in R-II, hence in (\ref{L_II}).  In this case, one expects two additional scaling parameters for the $\omega$, i.e., $g^\ast_\omega$ and $a^\ast_\omega$. In the analysis made below, we will see how these parameters are constrained by the EoS for compact stars.

\subsection{Effect on the Tensor Forces} \label{tensor_force}
In this subsection, we describe the structure of the tensor forces affected by the IDDs given in the previous subsection.  Here the $\omega$ meson turns out to affect little the symmetry energy factor $S$. To see the qualitative feature of the tensor force in medium,  we use the non-relativistic ($\frac{k^2}{m_N^{\ast\, 2}} \ll 1$) form of the tensor potential, valid in the region we are considering  as the in-medium nucleon mass stays heavy. The tensor potential\cite{c14,Machleidt:1989tm} is given by
\be
V^T_M \left( r \right) &=& S_M \frac{f_{NM}^{\ast\,2}}{4\pi} \tau_1\, \tau_2\, S_{12}{{\cal I}(m^\ast_{M} r)} \label{tensorM}\\
{{\cal I}(m^\ast_M r)}&\equiv& m_M^\ast
\left( \left[ \frac{1}{(m_M^\ast r)^3} + \frac{1}{(m_M^\ast r)^2} + \frac{1}{3m_M^\ast r} \right] e^{-m_M^\ast r} \right)\,,\label{radial}
\ee where $M=\pi,\,\rho$, $S_{\rho(\pi)} = +1(-1)$ and
\begin{equation}
S_{12} = 3 \frac{\left(\vec{\sigma}_1 \cdot \vec{r} \,\right)\left(\vec{\sigma}_2 \cdot \vec{r} \,\right) }{r^2} - \vec{\sigma}_1 \cdot \vec{\sigma}_2
\end{equation} with the Pauli matrices $\tau^i$ and $\sigma^i$ for the isospin and spin of the nucleons with $i = 1,2,3$.
The strength $f_{NM}^\ast$ scales as
\begin{equation}
R_M\equiv \frac{f_{NM}^\ast}{f_{NM}} \approx \frac{g_{M NN}^\ast}{g_{M NN}} \frac{m_N}{m_N^\ast} \frac{m_M^\ast}{m_M}
\end{equation}
where $g_{MNN}$ are the effective meson-nucleon couplings. Their relations to the coupling constants that figure in the Lagrangians will be specified below. What is significant in (\ref{tensorM}) is that given the same radial dependence, the two forces (through the pion and $\rho$ meson exchanges) come with an opposite sign and this well-known fact plays a crucial role.

First, we discuss the d-scalings of the tensor forces in medium given by IDDs and predict how the net tensor force scales in density. For the $\pi$ tensor force, applying  IDD$_{pNG}$(in R-I) and IDD$_{matter}$(dominantly in R-II) to $\pi$ and $N$, we have from Eqs.~(\ref{matter-mass}), (\ref{gpinn}) and (\ref{m-II})
\be
R_\pi &\approx& \frac{g_{\pi NN}^\ast}{g_{\pi NN}} \frac{m_N}{m_N^\ast} \frac{m_\pi^\ast}{m_\pi} \\
      &\approx& \left\{\begin{array}{ll}\Phi_I \times \Phi_I^{-1} \left(\frac{m_\pi^\ast}{m_\pi} \right) \quad \textnormal{for R-I}\label{Rpi-I}\\
                                        \kappa \times \kappa^{-1} \left(\frac{m_\pi^\ast}{m_\pi} \right) \quad \textnormal{for R-II} \end{array} \right.\label{Rpi-II}
\ee
hence
\be
   R_\pi\approx  \frac{m_\pi^\ast}{m_\pi}\quad \textnormal{in both R-I and R-II}.
\ee
 Thus the $\pi$-tensor force principally depends only on the d-scaling of $m_\pi^\ast$. It turns out as has been assumed since a long time that the pion tensor is insensitive to density:
 Within R-I, to the extent that the small pseudo-NG pion mass is in some sense protected by chiral symmetry, we expect the d-scaling of ${R_\pi}^2$ to be small. And so will be $ {{\cal I}(m^\ast_\pi r)}$. In addition the product of the former -- decreasing -- and the latter -- increasing -- largely cancels out. Thus the pion tensor does not d-scale in R-I.
  As for R-II, the situation is somewhat more involved. While $R_\pi$ is still expected to scale proportionally to the in-medium pion mass, the pion mass will not d-scale proportionally to $\sqrt{\la\bar{q}q\ra}$ since the bilinear quark condensate tends to zero for $n\gsim n_{1/2}$. The chiral symmetry is still spontaneously broken in R-II, hence we expect the GMOR relation, expected to hold in the tree (or mean-field) order medium, to be modified to{
 \begin{eqnarray}
&&f_\pi^{\ast\, 2} m_\pi^{\ast\, 2}
= m_q \langle \bar{q}q \rangle + \sum_{n > 1} c_n \langle \left(\bar{q}q\right)^n \rangle \\
&\Rightarrow& \kappa^2 f_{0\pi}^2 m_\pi^{\ast\, 2}
= \sum_{n > 1} c_n \langle \left(\bar{q}q\right)^n \rangle\, .
\end{eqnarray}}
We have indicated by the multiquark (or higher dimension filed) condensates possible non-vanishing contribution to the GMOR mass formula in which the quark condensate $\la\bar{q}q\ra$ is vanishing. There are several proposals for specific form for the order parameter(s)~\cite{multiquark-condensate}. Whatever the precise form may be, multiquark condensates are expected to be quite suppressed in R-II as shown in \cite{maetal}. This implies that the pion mass must decrease rapidly in R-II. Despite the rapid decrease of the pion mass in R-II,  the pion tensor remains non-d-scaling. This is shown in Fig.~\ref{pi_tensor}.

Now we turn to the d-scaling of the $\rho$-tensor force. At the mean-field (or tree) order in $s$HLS, the $\rho$-meson mass will satisfy the KSRF formula with IDD parameters
\begin{equation}
m^\ast_\rho = \sqrt{a^\ast_\rho}\, g^\ast_\rho\, f^\ast_\pi . \label{KSRF-medium}
\end{equation}
In the matter-free vacuum, the KSRF is a low-energy theorem proven to hold to all loop-orders in HLS~\cite{HKY-KSRF,HY:PR}.
Whether it also holds in medium to all loop-orders or not has not been proven. It seems however reasonable to assume that for a given density, this does hold with $g^\ast_\rho$  replaced by the effective $\rho NN$  coupling constant $g_{\rho NN}$. It has been shown in \cite{DLFP} that
\be
g^\ast_{\rho NN}=F^\ast_\rho g^\ast_\rho
\ee
with $F^\ast_\rho$ that goes to zero at the dilaton-limit fixed point (DLFP), possibly identical to the IR fixed point of \cite{CT}, independently of how $g^\ast_\rho$ d-scales. In our application in Section \ref{vlowk}, the effect of $F^\ast_\rho$ could in principle be included. Therefore we shall leave it out in what follows in our discussion, setting $F^\ast_\rho=1$, with the possibility in mind that $F^\ast_\rho$ effect could further speed up the dropping in $R_\rho$ given below.

While based on the d-scaling argument in \cite{LPR15} and relying on phenomenological observations~\cite{c14,BR:DD}, possible IDD$_{matter}$ effect in $g^\ast_\rho$ was ignored in R-I, the IDD$_{matter}$, as argued in Section \ref{region-II}, cannot be ignored in R-II.
%
%
%
%
Noting that
\begin{eqnarray}
m_\rho^\ast/m_\rho &\approx& \left(\frac{g^\ast_\rho}{g_\rho} \right) \left( \frac{f_\pi^\ast}{f_\pi}\right) \\
                   &\approx& \left\{\begin{array}{ll} \Phi_I \quad \textnormal{for R-I}\\
                                                      \Phi_{II}^\rho \times \kappa \quad \textnormal{for R-II} \end{array} \right.\,. \label{rhoM}
\end{eqnarray}
we find the d-scaling of $R_\rho$ to be of the form
\begin{eqnarray}
R_\rho &\approx& \frac{g_{\rho NN}^\ast}{g_{\rho NN}} \frac{m_N}{m_N^\ast} \frac{m_\rho^\ast}{m_\rho} \\
      &\approx& \left(\frac{g^\ast_\rho}{g_\rho} \right)^2 \\
      &\approx& \left\{\begin{array}{ll} 1 \quad \textnormal{for R-I}\\
                                         \left(\Phi_{II}^\rho \right)^2 \quad \textnormal{for R-II} \end{array} \right. \label{rhoR}
\end{eqnarray}

\begin{figure}[ht]
\begin{center}
\includegraphics[width=10.0cm]{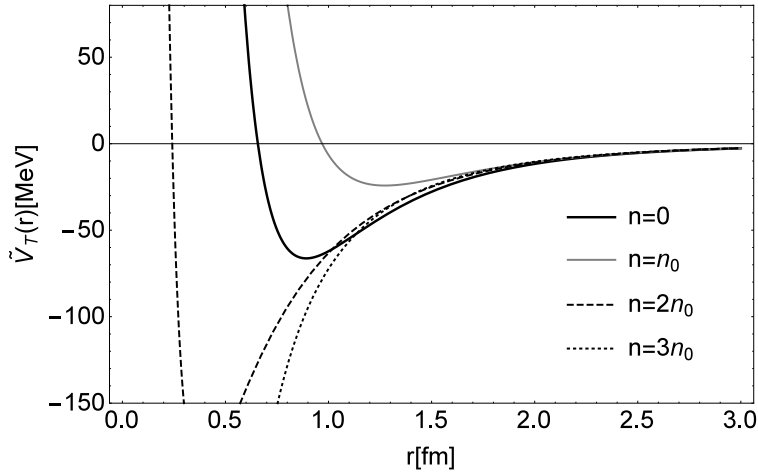}
\caption{$\tilde{V}_T\left(r\right) \equiv V_T\left(r\right) \left( \tau_1\, \tau_2\,S_{12}\right)^{-1}$. For illustration, we take $n_{1/2}=2n_0$, $\Phi_{I} = \Phi_{II}^\rho = 1 - 0.15 \frac{n}{n_0}$ and $\kappa = 1$. }
\label{tensorF}
\end{center}
\end{figure}

What is crucially important for the structure of the $\rho$ tensor force is the factor $R_\rho$ which changes discontinuously from R-I to R-II across $n_{1/2}$ with the topology change. In R-I, experimentally controlled nuclear processes indicate how $\Phi_I$ d-scales. It is a slow decrease reaching $\lsim 0.8$ at $n\sim n_0$. On the other hand, $\Phi_{II}^\rho$ is totally unknown. It is given neither theoretically nor phenomenologically. The only constraint based on HLS~\cite{HY:PR} is the vector manifestation fixed point at which $\Phi_{II}^\rho$ should approach 0 in the chiral limit. If the vector-manifestation (VM) fixed point is $\sim (6-7)n_0$ -- which is not too far from the density of the interior of $\sim 2$ solar-mass stars -- then $\Phi_{II}^\rho$ should drop more rapidly in R-II than in R-I. This point was already emphasized in \cite{BR:DD}. One can see from (\ref{rhoR}) that there will be  a rapid suppression of the $\rho$ tensor force at $n_{1/2}$.
This feature is shown in Fig.~\ref{tensorF}. Just for illustration, we have taken $\Phi_I=\Phi_{II}^\rho =1-0.15 \frac{n}{n_0}$ and $\kappa=1$.  In the application to the EoS for nuclear matter and compact-star matter, more realistic d-scaling of the parameters involved will be used.

{ It is now easy to see how the cusp structure in the $S$ factor at $n_{1/2}$ arises. In the density regime in the vicinity of nuclear matter,  the symmetry energy factor $S$, dominated by the tensor forces, can be reliably approximated by the closure formula~\cite{brown-machleidt},
\be
S\approx \frac{12}{\bar{E}}\la \bar{V}_T^2\ra \label{tensorSym}
\ee
where $\bar{E}\approx 200 {\rm MeV}$ is the average energy typical of the tensor force excitation and  $\bar{V}_T$ is the radial part of the net tensor force defined in Fig.~\ref{tensorF}. The  tensor force strength decreasing as density approaches  $n_{1/2}$ from below and increasing after above $n_{1/2}$ reproduces the cusp structure~\cite{LR tensor forces}. We suggest  that this feature provides, albeit indirect, support to the scaling properties formulated in a general term in \cite{LPR15}.

In Section \ref{eos}, the tensor force structure obtained above, together with the d-scaling properties in Regions I and II, will be confronted with the EoS of nuclear matter and neutron-star matter. For this, the renormalization group implemented $V_{lowk}$ technique will be employed. This will be briefly reviewed in Section \ref{vlowk}.
}

\section{Renormalization Group  with  $V_{lowk}$  }\label{vlowk}
As stated in Introduction, the scale-invariant HLS Lagrangian
with baryons ($bs$HLS) can be applied to many-nucleon systems in
either RMF involving single decimation or more microscopically with double decimations involving  $V_{lowk}$. Here we
briefly review the latter approach that will be used in
Section \ref{eos}. Details are found in the review articles referred in this paper. We will essentially follow the strategy
used in \cite{dongetal}.

One (in principle) starts with the NN potential $V_{NN}$ gotten from the
$bs$HLS Lagrangian (\ref{L_I}) and (\ref{L_II}) with the proper IDDs.
One then arrives at $V_{lowk}$ \`a la renormalization group technique\cite{bogner03,bogner03b}.
This consists of
 decimating the high momentum components from
the matching scale or $\Lambda_{eff}$ to a model-space  momentum scale
$\Lambda$ which is not far from the Fermi momentum $k_F$.
In terms of $T$-matrix,
this amounts to computing $V_{lowk}$ as
\be
T(k',k,k^2)=V_{\rm NN}(k',k)
+\frac{2}{\pi}\mathcal{P}\int_0^\infty
\frac{V_{\rm NN}(k',q)T(q,k,k^2)}{k^2-q^2}q^2dq,
\ee
\be
T_{\rm lowk}(k',k,k^2)=V_{\rm lowk}(k',k)
+\frac{2}{\pi}\mathcal{P}\int_0^\Lambda
\frac{V_{\rm lowk}(k',q)T_{\rm lowk}(q,k,k^2)}{k^2-q^2}q^2dq,
\ee
\be
 T(k',k,k^2)=T_{lowk}(k',k,k^2); (k',k) \leq \Lambda.
\ee
Here $\mathcal{P}$ denotes principal-value integration and the
intermediate state momentum \emph{q} is integrated from
0 to $\infty$ for the whole-space $T$ and from 0 to $\Lambda$
for $T_{\rm lowk}$.

With the given ``bare" effective Lagrangian,
if one wishes, one can do a sophisticated effective field theory
calculation (such as chiral perturbation theory) to obtain $V_{NN}$. This should be feasible starting with the effective Lagrangian we are dealing with, i.e., $bs$HLS.
For the exploratory work we are doing here, however, a rigorous EFT
calculation is unnecessary.
 In the present work, as in \cite{dongetal},  we choose the $V_{NN}$
to be the realistic BonnS \cite{bonns} NN interaction with the IDD
dependence encoded in the ``bare" parameters taken into account. We shall adopt the vacuum (free-space) parameters chosen in \cite{bonns} without adjustments.
This is a phenomenologically powerful approach, fit to experimental
data in free space as well as in medium to the momentum/energy
scale defined by the cutoff $\Lambda$.
Because we shall calculate the EoS, in particular, the nuclear symmetry energy, $E_{sym}(n)$ up to
$n\sim5n_0$, we shall use $\Lambda=3$ fm$^{-1}$ \cite{kuo}.
The $V_{lowk}$ so obtained preserves the low-energy phase shifts
 { in the vacuum} (up to energy $\Lambda^2$) and the deuteron
binding energy of $V_{NN}$. (For example, the deuteron binding energy
given by $V_{lowk}$ of $\Lambda=$ 2.0 and 3.0 fm$^{-1}$
are both -2.226 MeV.) By construction, $V_{lowk}$
is a smooth `tamed' potential which is suitable for being
used directly in many-body calculations.

 The first step in the procedure is to verify
 if in R-I the above  $V_{lowk}$ interaction can
satisfactorily reproduce  the empirical nuclear
matter saturation properties (saturation density
$n_0\simeq 0.16$ fm$^{-3}$ and average energy per nucleon
$E_0/A\simeq -16$ MeV at saturation). To do this, we shall
calculate $n_0$ and $E_0/A$ using a low-momentum ring-diagram
approach \cite{siu09,dong09,siu08,dong10,kuo}, where the $pphh$
ring diagrams are summed to all orders within a  model space of
the cutoff $\Lambda$.

\begin{figure}[h]
\begin{center}
\includegraphics[width=10.0cm]{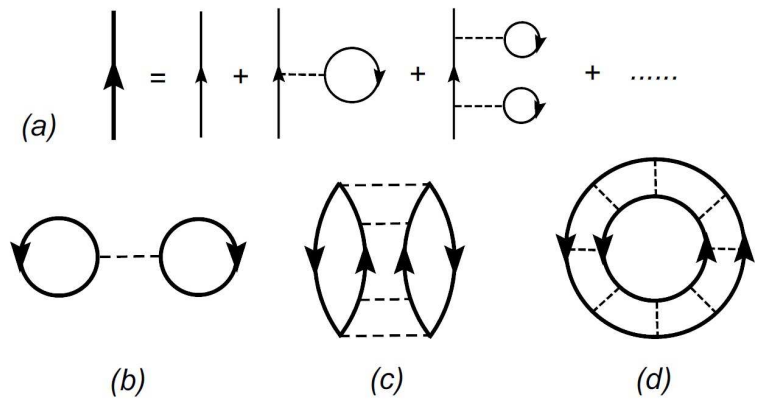}
\caption{Diagrams included in the $pphh$ ring-diagram summation
  for the ground state energy  of  nuclear matter. Included are
 self-energy insertions on the single-particle propagator as indicated
by (a), and  $pphh$ ring diagrams  by (b,c,d).}
\label{energy}
\end{center}
\end{figure}

We now briefly describe the above ring-diagram method.
The ground state energy shift is defined as
$\Delta E_0 = E_0-E_0^{\rm free}$
where $E_0$ is the true ground-state energy
 and the corresponding quantity for the non-interacting system
 $E_0^{\rm free}$. In the present work, we consider $\Delta E_0$
as given by the all-order sum of the  $pphh$ ring
diagrams as shown in (b), (c) and (d) of Fig.\ \ref{energy},
  they  being respectively
a (1st-, fourth- and eighth-order)  such ring diagram.
In our ring-diagram calculations, we also include  HF single-paticle
insertions to all orders as illustrated by (a) of the figure.
Note that each vertex
of the diagrams is a  $V_{lowk}$ interaction
  obtained from a  density-scaled  $V_{NN}$ potential.
We include in general three types of ring diagrams, the
proton-proton, neutron-neutron and proton-neutron ones.
 The proton and neutron
Fermi momenta are, respectively,
$k_{Fp}=(3\pi^2n_p)^{1/3}$
and $k_{Fn}=(3\pi^2n_n)^{1/3}$, where $n_p$ and $n_n$ denote respectively
the proton- and neutron-density. The asymmetric parameter is
$\alpha \equiv (n_n-n_p)/(n_n+n_p)$.
With such ring diagrams summed to all orders,  we have
\be\label{eng}
\Delta E(n,\alpha)=\int_0^1 d\lambda
\sum_m \sum_{ijkl<\Lambda}Y_m(ij,\lambda)  \times Y_m^*(kl,\lambda) \langle
ij|V_{\rm lowk}|kl \rangle,
\ee
where the transition amplitudes $Y$ are obtaind from a $pphh$ RPA equation
\cite{siu09,dong09}.
Note that $\lambda$ is a strength parameter,
integrated from 0 to 1. The above ring-diagram method reduces to the
usual HF method if only the first-order ring diagram
is included. In this case, the above energy shift becomes
$\Delta E(n,\alpha)_{HF}=\frac{1}{2}
\sum n_i n_j\langle ij|V_{\rm lowk}|ij \rangle$ where
$n_k$=(1,0) if $k(\leq,>)k_{Fp}$ for  proton
and $n_k$=(1,0) if $k(\leq,>)k_{Fn}$ for  neutron.

 The above $V_{lowk}$ ring-diagram framework has been applied to
symmetric and asymmetric nuclear matter \cite{siu09,dong09}
 and to the nuclear symmetry energy
\cite{kuo}.  This framework has also been tested by  applying it  to dilute cold neutron matter in the  limit that the $^1S_0$ scattering
length of the underlying interaction approaches infinity
\cite{siu08,dong10}. This limit -- which is a conformal fixed point --
 is usually referred to as the unitary limit, and the
corresponding potentials are the unitarity potentials.
 For many-body systems at this limit, the ratio
$\xi \equiv E_0/E_0^{free}$ is
expected to be a universal constant of value $\sim 0.44$.
($E_0$ and $
E_0^{free}$
have been defined earlier.)  The above ring-diagram method has been
used to calculate
neutron matter using several very different unitarity potentials
(a unitarity CDBonn potential  obtained by tuning
its meson parameters,
 and several square-well unitarity potentials) \cite{siu08,dong10}.
The $\xi$ ratios
given by our calculations for all these different unitarity potentials are
all close to 0.44, in good agreement with the Quantum-Monte-Carlo results
(see \cite{dong10} and references quoted therein).
In fact our ring-diagram results for $\xi$ are significantly better than
those given by HF and BHF (Brueckner HF) \cite{siu08,dong10}. It is
 desirable that the above unitary calculations have provided
satisfactory results,  supporting the reliability of
 our $V_{lowk}$ ring-diagram framework for calculating the nuclear matter EoSs.

 One should recognize that the above many-body approach is essentially equivalent to doing Landau Fermi-liquid fixed point theory with quasiparticle correlations on top of the Fermi sea treated with $V_{lowk}$ with IDDs implemented, as discussed in \cite{kuo-brown-fermiliquid}. This procedure is a microscopic improvement on the relativistic mean-field treatment involving single decimation of $bs$HLS Lagrangian. In the application to denser regime going into R-II, the above procedure will be simply extrapolated. It is most likely a valid procedure if the Fermi-liquid structure holds in R-II.

\section{EoS for Nuclear Matter and Compact-Star Matter}\label{eos}
In this section, we shall extrapolate the treatment
presented above, verified up to density $n_0$ and taken to be
valid up to $n_{1/2}\sim 2n_0$, to $n>n_{1/2}$ appropriate
to massive compact stars. {To calculate the EoS for nuclear matter, we use the Bonn A potential\cite{Machleidt:1989tm} consistent with the ``intrinsic density-dependent" $bs$HLS Lagrangian at the leading order of scale-chiral counting. Here, we should note that we fix the pion exchange potential not to scale in density for both R-I and R-II as we argued on the basis of the pion being nearly massless Nambu-Goldstone boson. This is a reasonable assumption for a qualitative account for the scaling of the parameters involved. As shown in \cite{Machleidt:1989tm}, the central, spin-spin and spin-orbit nuclear forces from one-pion-exchange are weak or negligible compared with the nuclear forces from the other particles while the tensor force from one-pion-exchange is strong. But, as we find in the appendix \ref{appen_tensor}, the pion-tensor is almost independent of a density.  }

We start with the ``bare'' parameters that figure in both R-I and
R-II. They are summarized in Table~\ref{summary}.
 \begin{table}
  \centering
  \begin{tabular}{c|c|}
     \hline \hline
  R-I & \begin{tabular}{c} $\frac{m^\ast_N}{m_N} \approx \frac{m^\ast_\sigma}{m_\sigma} \approx \frac{m^\ast_V}{m_V} \approx \frac{f^\ast_\sigma}{f_{0\sigma}} \approx \frac{f^\ast_\pi}{f_{0\pi}} \approx \Phi_I $\\ $\frac{g^\ast_\sigma}{g_\sigma} \approx \frac{g_V^\ast}{g_V} \approx 1 $\\ $\frac{g^\ast_{\pi NN}}{g_{\pi NN}} \approx \frac{g^\ast_A}{g_A} \approx \Phi_I \quad \& \quad \frac{m^\ast_\pi}{m_\pi} \approx  \left(\Phi_I \right)^{\frac{1}{2}}$\\ $\Phi_I = \frac{1}{1 + c_I \frac{n}{n_0}} $\end{tabular}\\ \hline
  R-II & $\begin{array}{ccc} \frac{m^\ast_N}{m_N} \approx \frac{m^\ast_\sigma}{m_\sigma} \approx \frac{f^\ast_\sigma}{f_{0\sigma}} \approx \frac{f^\ast_\pi}{f_{0\pi}} \approx \kappa \\ \frac{g^\ast_\sigma}{g_\sigma} \approx 1 \quad \& \quad \frac{g_V^\ast}{g_V} \approx \Phi^V_{II} \\
  \frac{g^\ast_{\pi NN}}{g_{\pi NN}} \approx \frac{g^\ast_A}{g_A} \approx \kappa \quad \& \quad m^{\ast\,2}_\pi \approx \frac{1}{f_{0\pi}^2\kappa^2} \sum c_n \langle \left( \bar{q}q \right)^n \rangle\\ \frac{m^\ast_\rho}{m_\rho} \approx \frac{g_\rho^\ast}{g_\rho} \quad \& \quad \frac{m^\ast_\omega}{m_\omega} \approx \kappa \sqrt{\frac{a^{\ast}_\omega}{a_\omega}}\frac{g_\omega^\ast}{g_\omega}\\
  \Phi^\rho_{II} = 1 - c_{II} \frac{n}{n_0} \quad \& \quad\Phi^\omega_{II}=  ?? \end{array}$\\ \hline \hline
  \end{tabular}
  \caption{The d-scaling ``bare" parameters of $bs$HLS Lagrangian  (\ref{L_I}) in R-I and (\ref{L_II}) in R-II. $\Phi^\omega_{II}$ is unknown if $U(2)$ symmetry is broken down as described in Appendix C. How it is deduced is discussed in the text.}\label{summary}
\end{table}
{As stated, there is only one constant, $c_I$, to be determined in R-I and two constants, $\kappa$ and $c_{II}$ in R-II. For the numerical work, we will adopt\footnote{The constant $c_I$ that figures in the double decimation $V_{lowk}$ approach needs not be the same as the single-decimation value found in the calculation of the anomalous orbital gyromagnetic ratio $\delta g_l$ measured in Pb~\cite{friman-rho}. In fact, it is found to be $c_I\approx 0.28$, somewhat larger than the range given in (\ref{numericalvalue}). There is no discrepancy here. The latter subsumes some part of quaiparticle interactions captured in the Landau fixed-point paramters $F_1$ and $F_2$.}
\be
c_I\approx 0.13 - 0.20, \ \ c_{II}\approx 0.15, \ \ \kappa\approx 0.7-0.8.\label{numericalvalue}
\ee
Before we confront Nature, we discuss how the formalism anchored solely -- with no fine-tuning -- on the three parameters of (\ref{numericalvalue}),  fares. We focus on the region R-I where there is only one parameter $c_I$ in the adopted parametrization $\Phi_I=1/(1+c_I\, n/n_0)$. In this region near $n_0$  fairly well-established experimental data are available.

First we recall that in R-I, the scaling is governed entirely by the d-scaling of $f^\ast_\sigma=\la\chi\ra^\ast$, i.e., the IDD$_{pNG}$. Although the explicit forms of the masses involved are not known in terms of the condensates, $\la\bar{q}q\ra$ and $\la G^2\ra$, we learn from CT theory~ \cite{CT} that the masses of nucleon and dilaton are dominated by $\la G^2\ra$ whereas the masses of the vector mesons are more crucially controlled by $\la\bar{q}q\ra$ in particular in R-II. This means that as the condensate $\la\bar{q}q\ra$ gets averaged to zero at $n_{1/2}$ with the possible multi-quark condensates nonzero but suppressed, the vector-meson masses will be more strongly affected by density than the masses of nucleon and dilaton. Thus one expects that $c_I$'s for $\rho$ and $\omega$ are larger than $c_I$'s for $\sigma$ and nucleon.  Now suppose we simply ignore this feature and take one universal $c_I$.  The result is given in Appendix \ref{AR-I}. One sees unambiguously that
 neither the equilibrium density nor the binding energy can be gotten with only $c_I$ parameter.  As we will see below, however, only a small adjustment within the given range of $c_I$'s, with the above scale symmetry feature taken into account, can reproduce fairly well the nuclear-matter observables. This exercise demonstrates that Nature seems to be  {\it unreasonably}  fine-tuned.

\begin{figure}[h]
\begin{center}
\scalebox{0.3}{
\includegraphics[angle=-90]{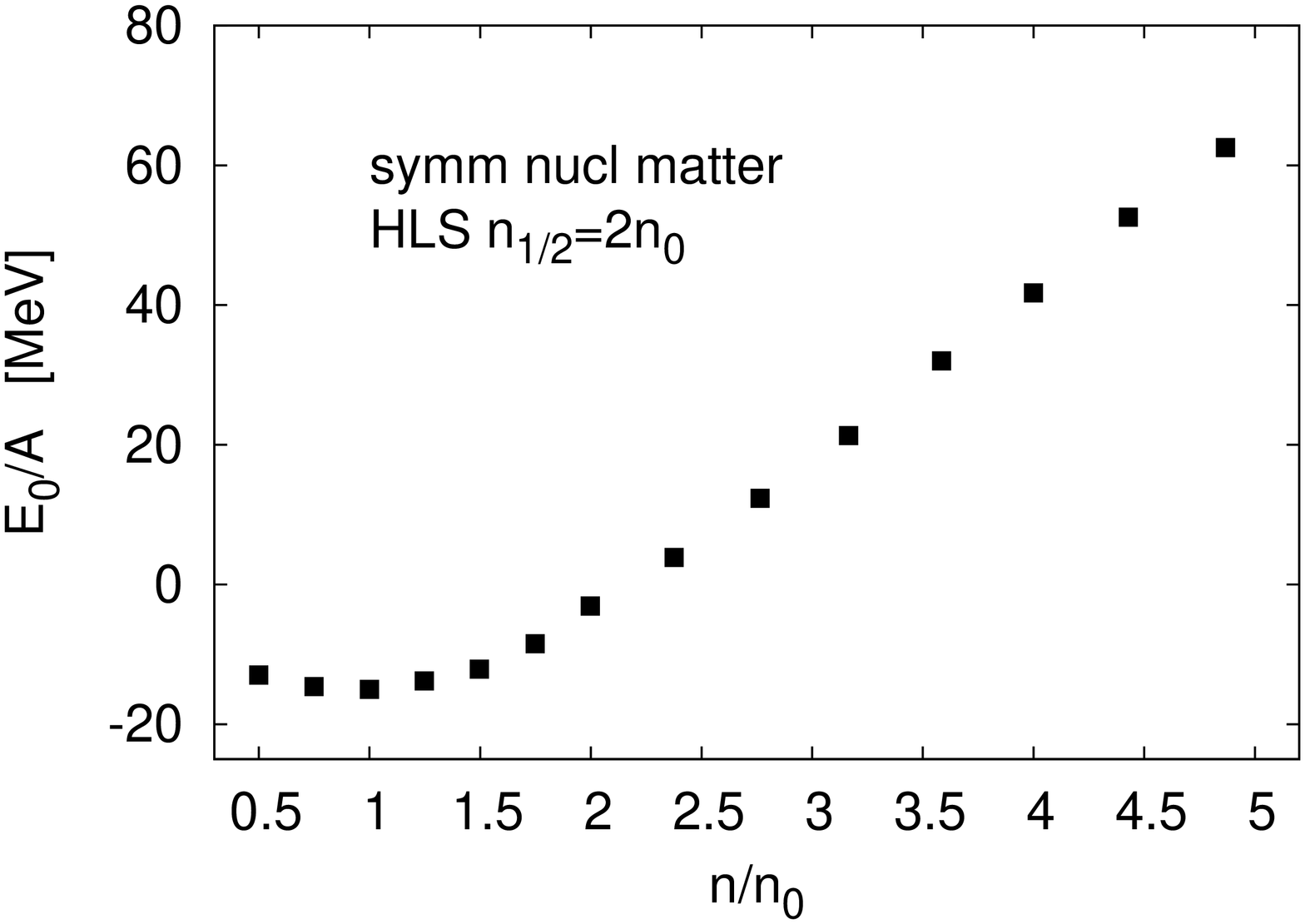}
\includegraphics[angle=-90]{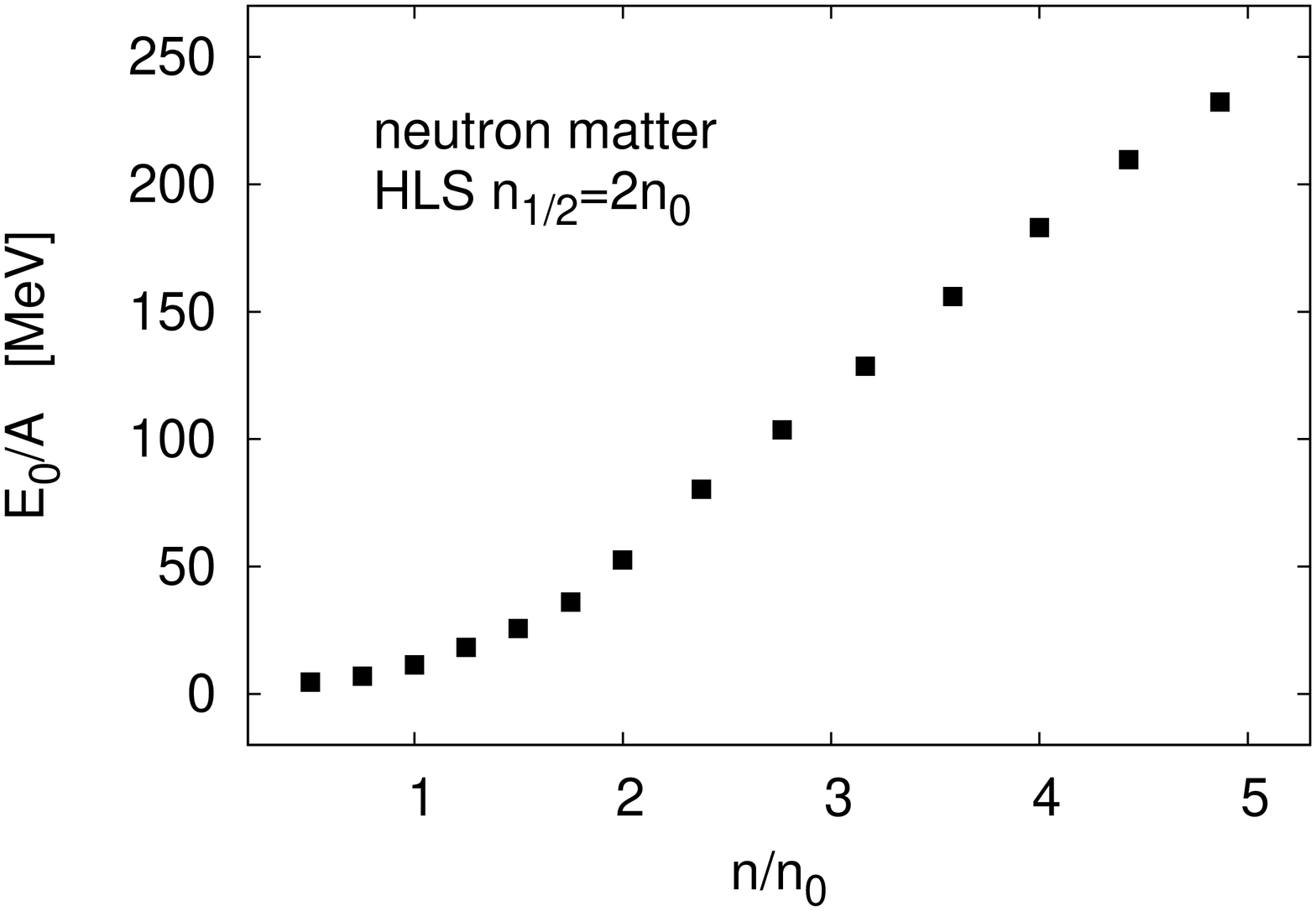}
}
\caption{Ground-state energy $E_0$ per nucleon
of symmetric nuclear matter (left panel) and neutron matter (right panel).
 }\label{E0}
 \end{center}
\end{figure}
Given that a universal $c_I$ fails quite dramatically to reproduce Nature, we  make, eschewing an extreme fine-tuning,  minimal adjustments for different mesons, within the range given in (\ref{numericalvalue}), to calculate  the ground-state energies of  nuclear matter using the ring-diagram method
with the density scaled $V_{lowk}$ interaction as described earlier. The small differences in $c_I$ may be considered as $1/N_c$ corrections in different channels in the ``bare" parameters of the in-medium Lagrangian.

Our results for symmetric nuclear matter and neutron matter are shown
respectively in the left and right panels of Figs.~\ref{E0}.
\begin{table}
  \centering
  \begin{tabular}{|c|c|c|}
     \hline
      & R-I &  R-II \\
     \hline
     $\frac{m_N^\ast}{m_N}$ & $\frac{1}{1+0.12\frac{n}{n_0}}$ & $0.71$ \\
     $\frac{m_\sigma^\ast}{m_\sigma}$ & $\frac{1}{1+0.12\frac{n}{n_0}}$ & $0.75$ \\
     $\frac{m_\rho^\ast}{m_\rho}$ &   $\frac{1}{1+0.14\frac{n}{n_0}}$ & $1 - 0.15 * \frac{n}{n_0}$ \\
     $\frac{m_\omega^\ast}{m_\omega}$ & $\frac{1}{1+0.14\frac{n}{n_0}}$ & $0.73\, \frac{g_{\omega}^\ast}{g_{\omega}}$ \\
     $\frac{g_{\rho}^\ast}{g_{\rho}}$ &   $1$ &$1 - 0.15 *\frac{n}{n_0} $ \\
     $\frac{g_{\omega}^\ast}{g_{\omega}}$ &  $1$ & $1 - 0.053 *\frac{n-n_{1/2}}{n_0} $ \\
     \hline
   \end{tabular}
  \caption{The precise values for the scaling parameters in R-I and R-II. The scaling properties shown above are consistent with the scaling of the parameters in Table \ref{summary}. {As stated in the text, the pionic parameters are taken not to scale in both R-I and R-II.} }\label{scalingpara}
\end{table}
{The scaling parameters employed are shown in Table \ref{scalingpara}.

For R-I, as indicated in Fig.~\ref{E0}, we determine $c_{I}$'s to
provide a satisfactory description of the saturation properties of symmetric nuclear matter, giving saturation energy
$E_0/A$=-15.1 MeV, the saturation density $n_{sat}=0.16$ fm$^{-3}$
and the compression modulus $K=$ 183.2 MeV.  The compression modulus comes out somewhat smaller than the value often quoted $\gsim 200$ MeV. This approach predicts a softer EoS for nuclear matter than one for neutron matter (given below). We will return to this matter ater. Table \ref{scalingpara} shows that the scaling in R-I is consistent with the expectation that $c_I$'s for $\rho$ and $\omega$ should be lager than $c_I$'s for $\sigma$ and nucleon.
It is to be noted that the inequality $c_I^{N,\sigma}< c_{I}^{\rho,\omega}$ argued for in the context of chiral-scale symmetry is crucial for the fit to Nature. Thus Nature seems to exercise a fine-tuning that goes beyond the general framework adopted in our approach for the EoS considered for compact stars. The coupling constants $g_{\rho}^\ast$  and $g_{\omega}^\ast$, associated with IDD$_{matter}$  do not scale in this region.

The scaling in R-II is our main interest. It will be shown below that
the scaling in Table \ref{scalingpara}, qualitatively consistent to that of  Table \ref{summary}, produces the  EoS for nuclear matter, which is compatible with Nature. For R-II, we take $\Phi_{II}^\rho = 1-0.15(n/n_0)$ for the VM behavior of the $\rho$ with $n_c \approx (6-7) n_0$ and $m^\ast_\rho/m_\rho = g^\ast_\rho/g_\rho = \Phi_{II}^\rho$. The oversimplified  parametrization for $m^\ast_\rho$ and $g^\ast_\rho$ may be the cause of the most likely artificial gap in the ground state energy at $n=n_{1/2}$. To remove this gap and get the resulting EoS within the empirical constraint of Danielewicz\cite{daniel02}, we adjust the values for $\kappa$'s of $\sigma$, $\omega$ and nucleon. As for the $\omega$ properties, we take a scaling drastically different from that of the VM behavior of the $\rho$, say, $\Phi_{II}^\omega = 1-0.053(n-n_{1/2})/n_0$. Given that the attraction is largely controlled by the dilaton exchange whose mass remains unscaling or at most slowly scaling, the repulsion due to the $\omega$ exchange is highly constrained, so that a faster decrease of the $g^\ast_{\omega}/g_{\omega}$ cannot be accommodated. Likewise a substantial decrease of $a^\ast_{\omega}$ would not be allowed if one were not to exceed the Danielewicz constraint. Therefore we have simply taken $a^\ast_{\omega}=a_{\omega}$.}

\begin{figure}[h]\begin{center}
\scalebox{0.4}{
\includegraphics[angle=-90]{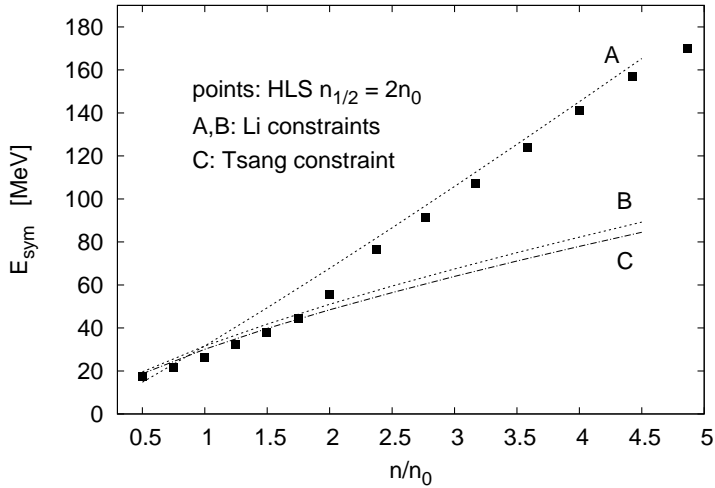}
}
\caption{ Comparison of our calculated
symmetry energies $E_{sym}(n)$ with the empirical ones
of Li $et$ $al.$ \cite{li05} and Tsang $et$ $al.$ \cite{tsang09}. It is worth noting that the predicted symmetry energy manifests a shift from soft to hard at $n_{1/2}$ reflecting the classical cusp structure in the skyrmion description of the topology change.}\label{sym-energy}
\end{center}
\end{figure}
As indicated by Eq.(\ref{E}), the nuclear symmetry energy $E_{sym}(n)$
is given by $E_0(n,1)/A-E_0(n,0)/A$. (Here $E_{sym}$ is the same as the
S factor of Eq.(\ref{E}).) In Fig.~\ref{sym-energy} we present
our calculated symmetry energies, and compare
them with their empirical values.  Li \emph{et al.} \cite{li05}
have suggested an empirical relation
\begin{equation}
E_{sym}(n) \approx 31.6 {\rm MeV}(n/n_0)^\gamma;~ \gamma=0.69-1.1,
\end{equation}
for constraining the density dependence of the symmetry energy.
The upper ($\gamma = 1.1$) and lower ($\gamma=0.69$) constraints are also
plotted in the figure, labelled as A and B respectively.
Tsang \emph{et al.} \cite{tsang09}
 proposed an empirical formula for the symmetry energy, namely
\begin{equation}
E_{sym}(n)=\frac{C_{s,k}}{2}\left(\frac{n}{n_0}\right)^{2/3}+
\frac{C_{s,p}}{2}\left(\frac{n}{n_0}\right)^{\gamma_i}
\end{equation}
where $C_{s,k}=25\, {\rm MeV}$, $C_{s,p}=35.2\, {\rm MeV}$
and  $\gamma_i \approx 0.7$. This formula is also plotted
in Fig.~\ref{sym-energy}, labelled as C. As seen, our calculated
$E_{sym}$ is slightly lower than the constraints in the low-density
region, and is close to Li's upper constraint in the high-density
region. It is noteworthy that the $S$ is relatively soft at low densities and hard at high densities, the changeover occurring at the crossover density $n_{1/2}$. This is a prediction of our theory.

Extensive studies have been carried out by Lattimer and Lim
\cite{lattimer12}  concerning the empirical constraints
on $E_{sym}$ and
 $L$ (defined as $ 3u(dE_{sym}/du) , u\equiv n/n_0$)
at density $n=n_0$. Their results deduced from a wide range
of observables including nuclear masses, nuclear giant
dipole resonances, astrophysics and neutron skins of the $Sn$ isotopes
suggested
$ 28 \lsim E_{sym}/{\rm MeV} \lsim 32 $ and
$  40 \lsim L/{\rm MeV}  \lsim 60$.
Our results for them are given in  Table \ref{tablesym}, indicated by '$bs$HLS'.  Our
 $E_{sym}$ is $\sim 27$ MeV which is slightly smaller than the
empirical value of $\sim 30$ MeV. Our $L$ value of $\sim 57$ MeV
is in satisfactory agreement
with the empirical values of Lattimer, but slightly lower than Li's lower and Tsang's constraints. (The two entries in row 2 of the Table \ref{tablesym} are respectively
the $L$ values given by Li's lower and upper constraints.)

\begin{table}[ht]
\centering
\begin{tabular}{c|c|c}\hline \hline
 $E_{sym}/MeV$ & $L/MeV$         \\ \hline
 27   &   57.3  & $bs$HLS \\
 31.6   &   65.4-104.2 & Li \\
 30.1   &   62.0  &  Tsang \\
 28-32  &   40-60 & Lattimer \\
\hline \hline
\end{tabular}
\caption{Comparison of our calculated $E_{sym}$ and $L$ (bsHLS)
at $n=n_0$ with the empirical values of Li et al. \cite{li05}, Tsang et al.
\cite{tsang09} and Lattimer et al \cite{lattimer12}. } \label{tablesym}
\end{table}

\begin{figure}[h]
\begin{center}
\scalebox{0.3}{
\includegraphics[angle=-90]{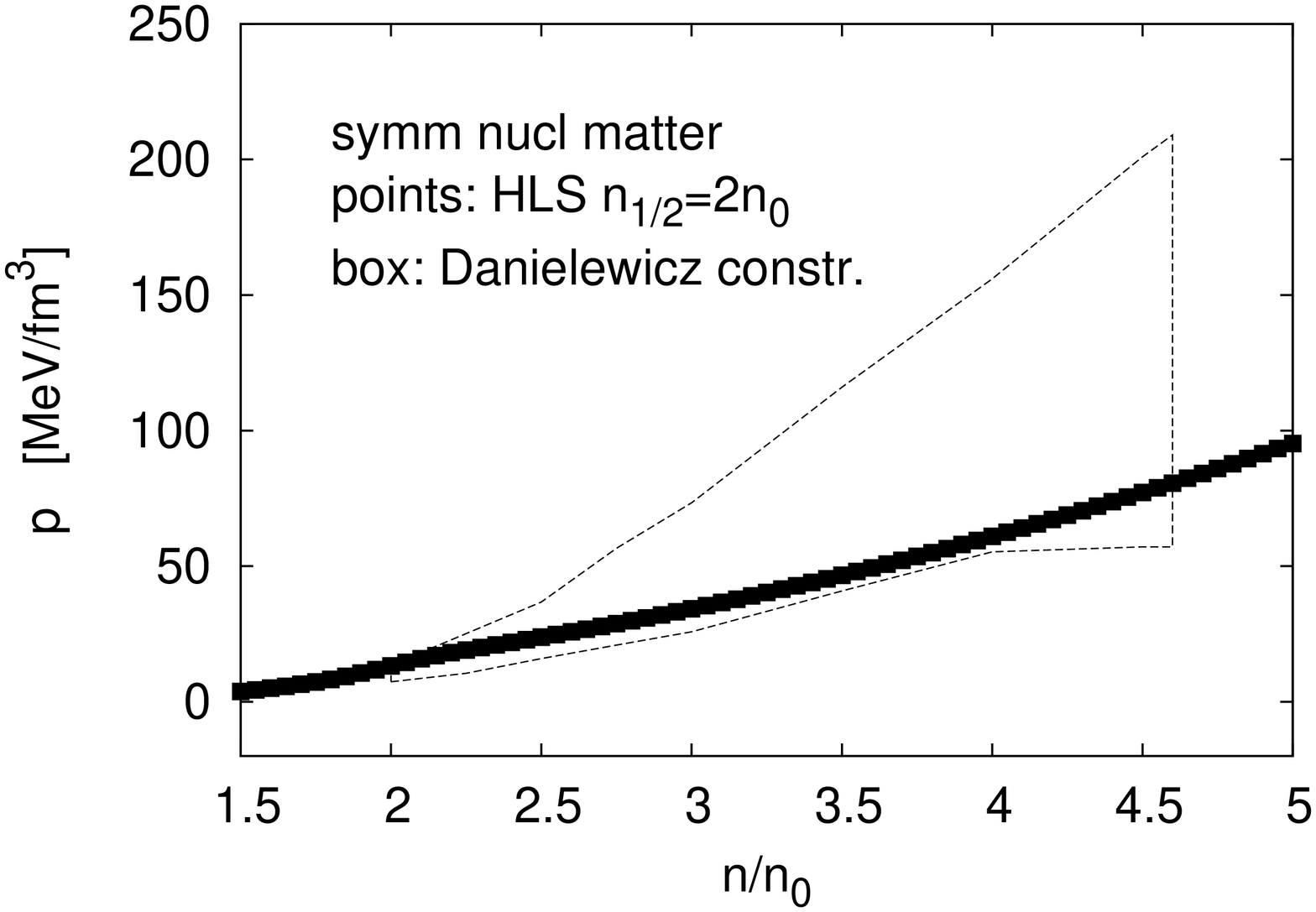}
\includegraphics[angle=-90]{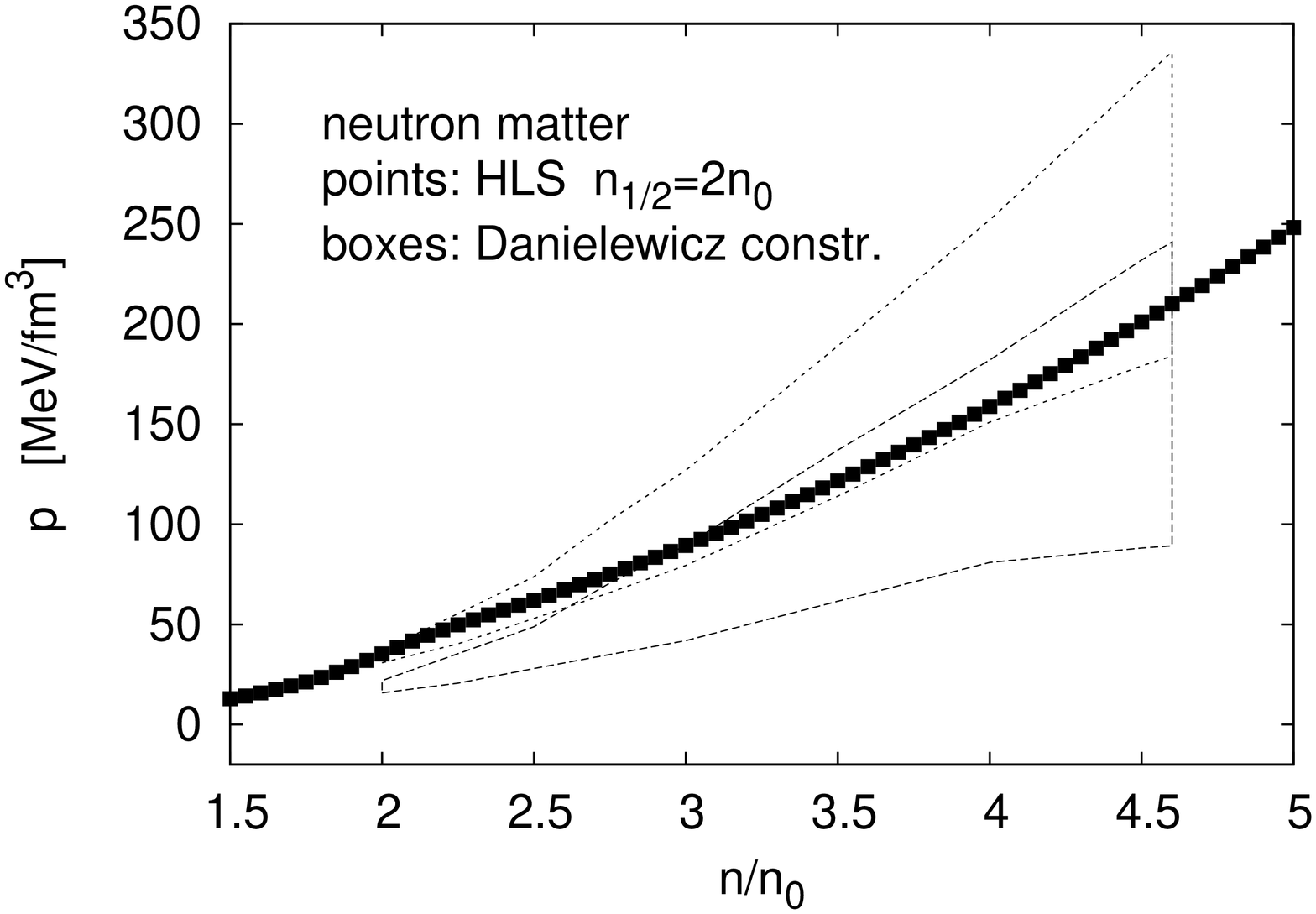}
}
\caption{ Calculated
pressure in symmetric nuclear matter (left panel) and the same
in neutron matter (right panel) compared with the Danielewicz
constraints \cite{daniel02}. }\label{pressuredan}
\end{center}
\end{figure}
It is of interest and useful to calculate the pressure-density EoS
$p(n)$ and compare it with the empirical constraints of
Danielewicz $et ~al.$ \cite{daniel02}. This EoS is given by using
\begin{equation}
\frac{E_0}{A} = a\left(\frac{n}{n_0}\right) + b \left(\frac{n}{n_0}\right)^c
\end{equation} to fit $E_0/A$ in Fig. \ref{E0}, where
\begin{equation}
p (n)=n\frac{d\epsilon(n)}{dn} - \epsilon(n)
\end{equation} and the energy density is
\begin{equation}
\epsilon (n)=n[\frac{E_0(n)}{A}+m_N]
\end{equation}
with $m_N$ being the nucleon rest mass.
Our result for $p(n)$  of  symmetric
 nuclear matter is shown in the left panel of Fig.~\ref{pressuredan}.
Our EoS is generally in agreement with the Danielewicz constraint,
although being rather close to the lower boundary of the constraint box.

\begin{figure}[h]
\begin{center}
\scalebox{0.34}{
\includegraphics[angle=-90]{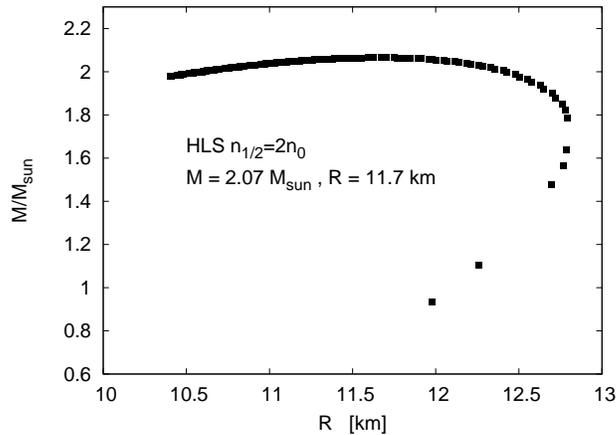}
}
\caption{Mass-radius relation of {pure neutron stars} calculated
from the EoS of Fig.~\ref{pressuredan} (right panel). Our EoS does not contain the part of EoS of low densities appropriate for the surface region, hence cannot account for the low-mass stars.}\label{M-R}
\end{center}
\end{figure}

\begin{figure}[h]
\begin{center}
\scalebox{0.34}{
\includegraphics[angle=-90]{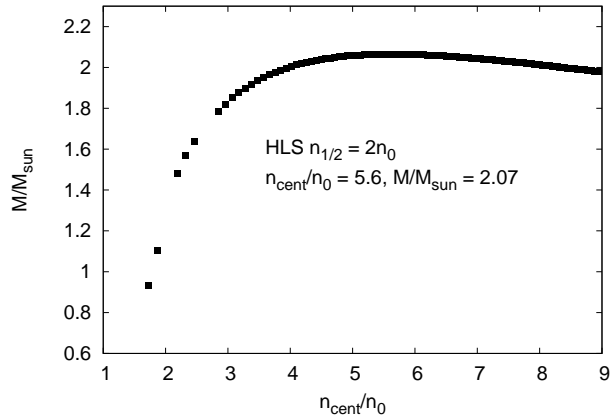}
}
\caption{Central densities of pure neutron stars  of Fig.~\ref{M-R}.}\label{M-n}
\end{center}
\end{figure}

Our calculated $p(n)$ for neutron matter is shown in
right panel of Fig.~\ref{pressuredan}, and
as shown it is generally in agreement with the
Danielewicz constraints.
 There are two Danielewicz
constraints in this case, one for
 the empirical stiff EoS (upper box) and the other
for the soft one (lower box).
Our $p(n)$, being near the lower boundary of the upper (stiff) box,
is mostly in the overlapping region allowed by both constraints.

Our neutron-matter EoS may be considered as `medium stiff'.
Can it support a massive neutron star such as the $2M_{\odot}$ one
recently observed by Antoniadis et al.~\cite{antoniadis13}?
From the neutron matter
EoS we can calculate the properties of pure neutron stars. This is
of much interest, and could provide a  useful test of our
neutron matter EoS in the high density region.
We have done so and our results are presented
below. We first calculate the pressure-energy EoS $p(\epsilon)$
and then  various neutron-star properties are obtained
from solving the Tolman-Volkov-Oppenheimer (TOV)
equations with the above EoS as the input. (See e.g. \cite{dong09}).
\begin{figure}[h]
\begin{center}
\includegraphics[width=8cm,angle=-90]{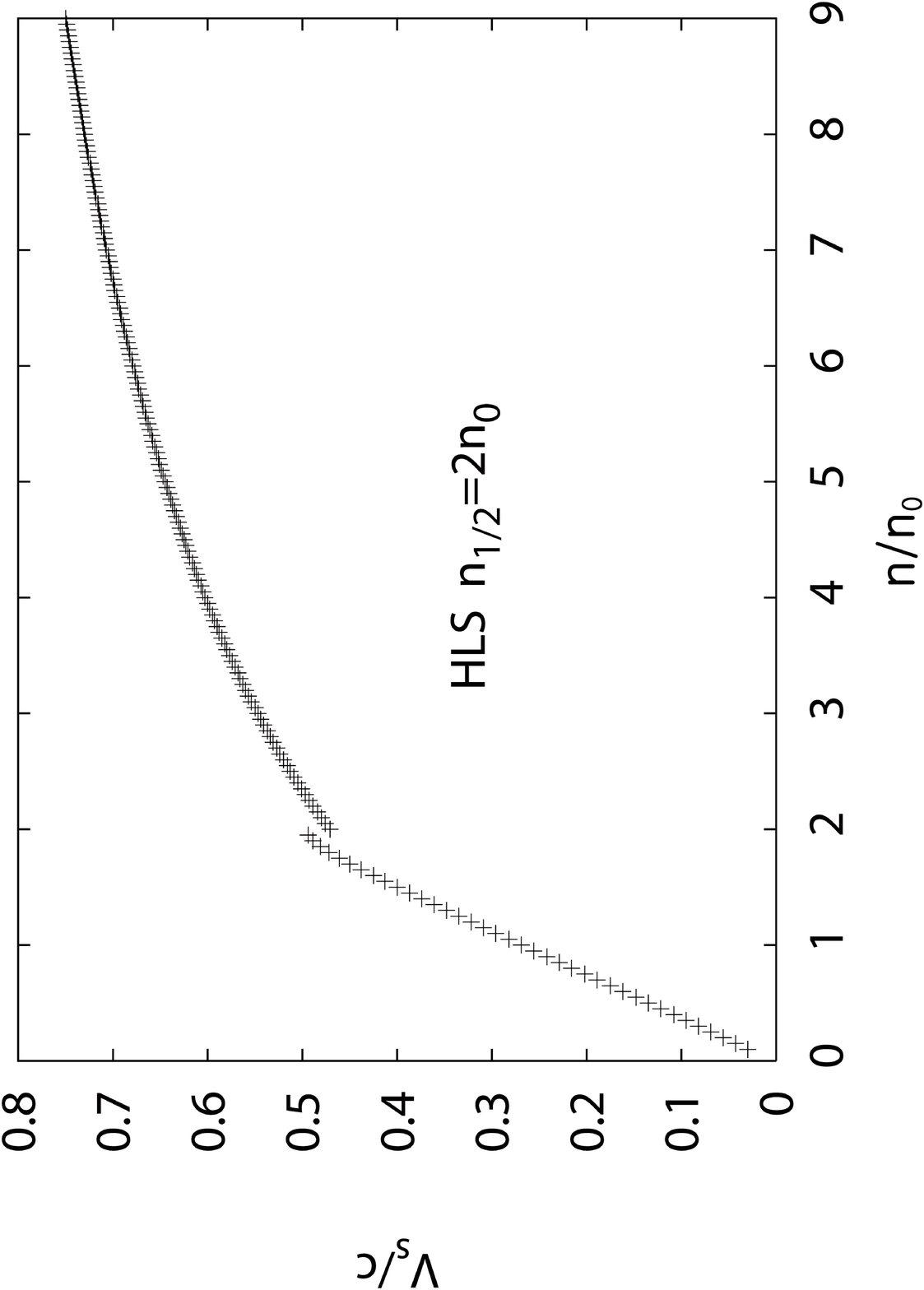}
\caption{ The sound velocity obtained from a fitting formula that reproduces the EoS of the neutron matter in Fig.~\ref{E0}.  Near the crossover density $n_{1/2}= 2n_0$, there is some discontinuity -- most likely artificial -- in velocity caused by the sharp demarcation of the parameter scaling, which turns out to be sensitive to the way the fitting is done.}
\label{pressure}
\end{center}
\end{figure}
\begin{figure}[h]
\begin{center}
\scalebox{0.4}{
\includegraphics[angle=-90]{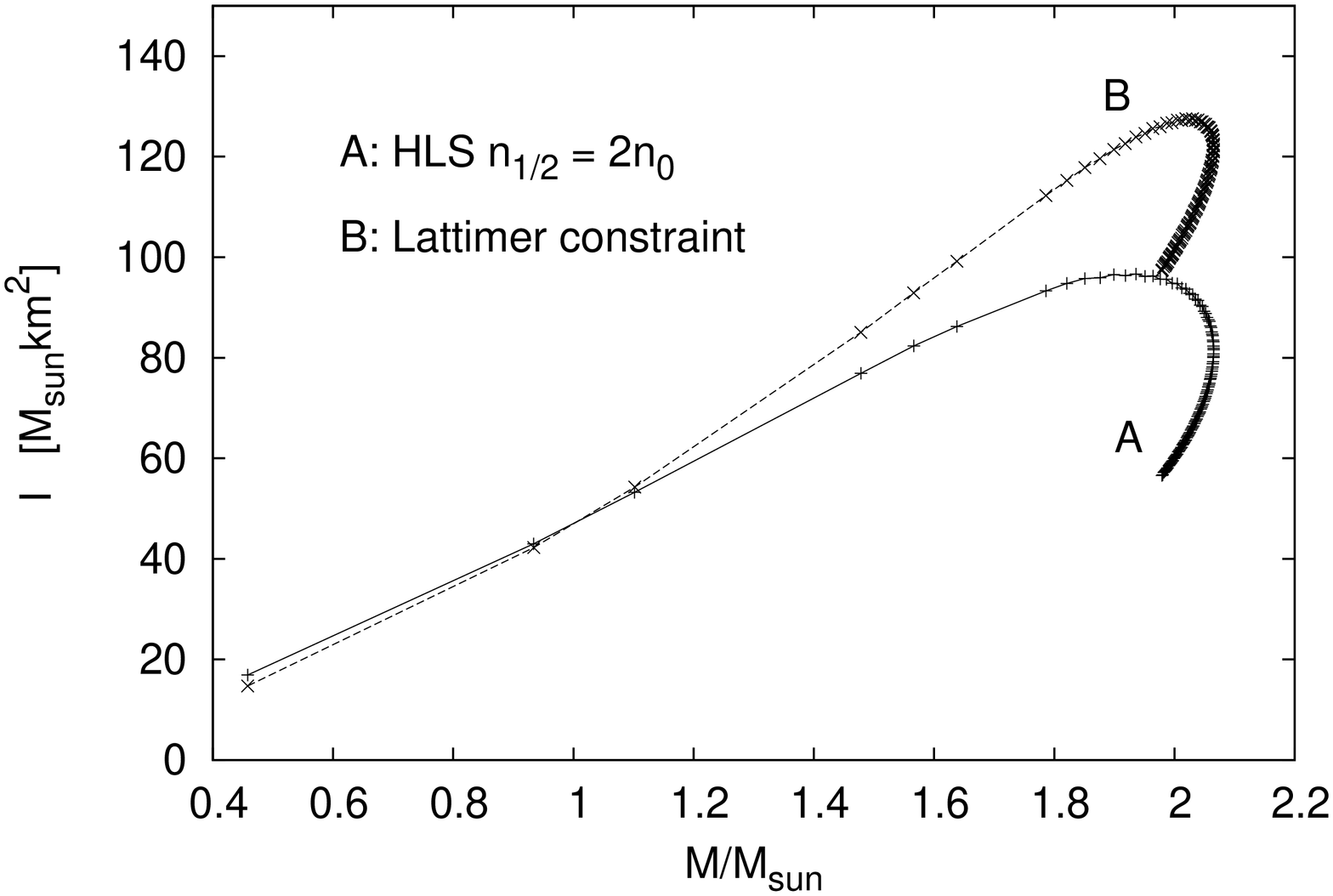}
}
\caption{Moment of inertia of neutron stars
 of Fig.~\ref{M-R}.  }\label{mom-inertia}
 \end{center}
\end{figure}

 In Fig.~\ref{M-R} our calculated neutron-star mass-radius relation
is shown. Our maximum-mass neutron star has mass $M \simeq 2.07 M_{\odot}$,
and radius  $R\simeq 11.7$km. With the weak equilibrium, not included here, taken into account, we expect that the maximum mass will come down a bit.
This calculated mass  is close
to the mass $2.01 \pm 0.04 M_{\odot}$ of the recently
observed relativistic pulsar~\cite{antoniadis13}.
  It may be worth mentioning that the central density of our
maximum-mass neutron star is merely $\sim 5.6 n_0$
as indicated in Fig.~\ref{M-n}. At this density, we have found that
our EoS is within the causual limit; for example
our EoS has $v_{sound}/c \sim 0.65$ at densities between
$n/n_0 =$ 5.0 and 6.0. This is shown in Fig.~\ref{pressure}.
What is significant in this result is that the change in the d-scaling in the parameters of the effective Lagrangian in R-II stabilize the sound velocity within the causal limit. Without the topology change, the causality would be violated at the density reached in massive stars.

Lattimer and Schutz have proposed an empirical
relation
\begin{equation}
I \simeq (0.237 \pm 0.008)MR^2
[1+4.2 \frac{M}{M_{\odot}}\frac{km}{R}
  + 90(\frac{M}{M_{\odot}}\frac{km}{R})^4]
\end{equation}
constraining $M$, $R$ and the moment of inertia $I$
of neutron stars~\cite{lattimer05}.
In solving the TOV equations, we integrate outward from the center of
the neutron star till its edge where pressure is zero. In this process
we know the matter distribution at all radii and thus can calculate
its moment of inertia I. In Fig.~\ref{mom-inertia}, we compare this I (A)
with the one given by the above relation using our calculated
$M$ and $R$ as inputs (B). This comparison provides a check
of our calculated density profile of the neutron star.
As seen, our results are in good agreement with the empirical
values for neutron stars with mass less than $\sim 1.4 M_{\odot}$
But at larger masses, there is significant difference
between the two. As of now, we are not sure about the reason for
this difference. In our present calculation, neutron stars
are assumed to be composed of pure neutron matter. Maybe the
above difference is related to this assumption; namely this
assumption is possibly adequate for light neutron stars but
not so for heavier ones. For neutron stars of mass larger than
$\sim 1.4M_{\odot}$,
the presence of other constituents such as protons~\cite{dong09}
and/or strange particles may be necessary. In the following section,
we consider  the latter possibility.

\section{Strangeness Problems}\label{hyperon}
We discuss in this section how the approach formulated in flavor $SU(2)$ exploited above for the EoS of compact stars can be applied to strangeness in compact-star matter, namely, the hyperon puzzle and kaon condensation problem. In the treatment given above, the effect of the dilaton which has a natural habitat in flavor $SU(3)$ is projected into the $SU(2)$ HLS Lagrangian.  To be fully realistic in addressing strangeness, one should resort to 3-flavor scale-invariant HLS Lagrangian with the dilaton treated on the same footing with kaons~\cite{CT}.  To address the EoS of compact stars with strangeness duly taken into account, the $V_{lowk}$ RG approach would then have to be extended to three flavors. Unfortunately such a three-flavor $V_{lowk}$ formulation is not yet available. In this section, we sketch how to address some of the issues involved with strangeness in the two-flavor framework developed applied in Section \ref{eos}.
\subsection{Hyperon problem}\label{hyperon}
In developing the EoS in Section \ref{eos}, strangeness degrees of freedom have been ignored. It is known that if the strangeness enters in compact-star systems, then the EoS can become too soft and massive stars of $\sim 2$ solar mass cannot be supported. For instance, this applies to the presence of hyperons. Simple energetic considerations suggest that hyperons should be present at high density in compact-star matter~\cite{glendenning}. The lowest-lying hyperon $\Lambda$, with its attractive interaction, is estimated to appear at matter density $\sim 2n_0$ with the others possibly appearing at higher density. What this suggests is that the hyperons could appear at about the same density as the one at which the half-skyrmion phase appears in the skyrmion matter. If this were the case, then the prediction made in Section \ref{eos} would make no sense without the hyperonic degree of freedom taken into account.

 Since a $V_{lowk}$ formalism for 3-flavor is not available, we address this problem using a mean-field approach with the two-flavor $bs$HLS Lagrangian employed in Section \ref{eos}. One can think of this approach as a ``single-decimation" RG approach as introduced in \cite{BR:DD} in contrast to the double-decimation applied above. This approach was applied with success to the calculation of the anomalous gyromagnetic ratio in heavy nuclei $\delta g_l$~\cite{friman-rho}.

We find that with the scaling property of the ``bare" parameters of the Lagrangian obtained above, the interactions between $\Lambda$s and nucleons become sufficiently repulsive at a density $n\lsim 3n_0$ so as to prevent the $\Lambda$s from appearing in the system. Our reasoning relies on what we shall refer to as ``Bedaque-Steiner" constraint

In a highly sophisticated phenomenological study using a Monte Carlo simulation over parameters that enter in the EoS for symmetric and asymmetric nuclear matter such as the compression modulus $K$ and $L$, symmetry energies $S$ and $S_\Lambda$, Bedaque and Steiner obtain the range of density $\Delta$ constrained by hydrodynamic stability of the system, that ensures that stars with $M >2M_\odot$ could be supported~\cite{bedaque-steiner}.  The $\Delta$ is then the range of density beyond which the in-medium $\Lambda$ mass becomes greater than the vacuum value. One expects -- and it is confirmed experimentally -- that the $\Lambda$-nucleon interaction is attractive at normal nuclear matter density, so $\Lambda$s can be bound in nuclear matter. In compact star matter, as density increases, the chemical potential difference between neutron and proton increases and it can become energetically favored to have spontaneous creating of hyperons in the system. It turns out that this can happen when density reaches roughly twice the normal nuclear matter density unless the attractive interaction is cancelled by repulsive mechanisms. The instability generated by the presence of hyperons at that low density is the hyperon problem. The $\Delta$ then stands for the range of density at which the $\Lambda$-interactions must be repulsive enough to make the in-medium $\Lambda$ mass be greater than the vacuum mass. The analysis by Bedaque and Steiner establishes that  the range of $\Delta$ required is $1\lsim \Delta/n_0\lsim3$.

\begin{figure}[!t]
\begin{center}
\includegraphics[width=0.4\textwidth]{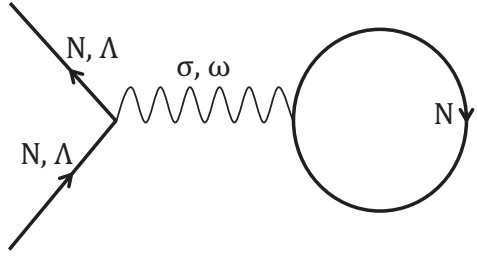}
\caption{Tadpole diagram for self-energies for the nucleon $N$ and the hyperon $\Lambda$ in medium. The loop corresponds to the nucleon scalar density $n_s=\la\bar{N}N\ra$ for coupling to $\sigma$ and the nucleon number density $n=\la N^\dagger N\ra$ for coupling to $\omega$.}
\label{tadpole}
\end{center}
\end{figure}

It is feasible, with some reasonable assumptions,  to calculate the effective mass of $\Lambda$ in medium using the $bs$ HLS formalism applied above. We do this using an RMF approximation with the Lagrangian (\ref{L_I}) and (\ref{L_II}).

In the mean field approximation, the chemical potential for the $\Lambda$ in medium gets contributions from two sources, one from the IDD$_{pNG}$ in the ``bare'' mass parameter $m^*_\Lambda\approx \frac{f^\ast_\sigma}{f_{0\sigma}} m_\Lambda$ and the other from the potential terms coming from $\Lambda$-nuclear coupling via $\sigma$ and $\omega$ exchanges as depicted in Fig.~\ref{tadpole},
\begin{eqnarray}
\mu_\Lambda
= m^*_\Lambda -\frac{g^\ast_{\sigma\Lambda}g^\ast_{\sigma N}}{m^{\ast\,2}_{\sigma}} n_s +\frac{g^\ast_{\omega \Lambda}g^\ast_{\omega N}}{m^{\ast\,2}_{\omega}} n\, \label{muL2}
\end{eqnarray}
where $n_s$ and $n$ are, respectively, nucleon scalar density and nucleon number density as defined in Fig.~\ref{tadpole}. The notations for $(\sigma,\omega)$ coupling to $\Lambda$ and $N$ are self-evident. The asterisk stands for d-scaling parameters.

\begin{table}
  \centering
  \begin{tabular}{c|c|}
     \hline
      & Parameters  for $\Lambda$ Mass Shift   \\
     \hline
     \hline
          & $\frac{m^\ast_{\Lambda}}{m_{\Lambda}} = \frac{m^\ast_{N}}{m_{N}} = \frac{m^\ast_{\sigma}}{m_{\sigma}} = \frac{m^\ast_{\omega}}{m_{\omega}} = \frac{1}{1+c_I* \frac{n}{n_0}}$\\
     R-I & $\frac{g^\ast_{\omega \Lambda}}{g_{\omega \Lambda}} = \frac{g^\ast_{\omega N}}{g_{\omega N}} = \frac{g^\ast_{\omega}}{g_{\omega}} = \frac{g^\ast_{\sigma \Lambda}}{g_{\sigma \Lambda}} = \frac{g^\ast_{\sigma N}}{g_{\sigma N}} = 1$\\
     \hline
          & $\frac{m^\ast_{\Lambda}}{m_{\Lambda}} = \frac{m^\ast_{N}}{m_{N}} = \frac{m^\ast_{\sigma}}{m_{\sigma}} = \kappa =\frac{1}{1+c_I* \frac{n_{1/2}}{n_0}}$\\
     R-II & $\frac{m^\ast_{\omega}}{m_{\omega}} = \kappa \frac{g^\ast_{\omega}}{g_{\omega}}$ $\,\&\,$ $\frac{g^\ast_{\omega \Lambda}}{g_{\omega \Lambda}} = \frac{g^\ast_{\omega N}}{g_{\omega N}} = \frac{g^\ast_{\omega}}{g_{\omega}}$ \\
          & $\frac{g^\ast_{\sigma \Lambda}}{g_{\sigma \Lambda}} = \frac{g^\ast_{\sigma N}}{g_{\sigma N}} = 1$\\
     \hline
     \hline
   \end{tabular}
  \caption{The ``bare'' parameter scaling for mean-field estimate of $\Lambda$ mass shift in dense matter. The only scaling parameter is chosen to be $c_I=0.13$ as in Section \ref{eos}.  The vacuum scalar (dilaton) mass is taken to be $m_\sigma=720\,{\rm MeV}$ so as to give $\sim 600$ MeV at nuclear matter density appropriate for RMF approach.  We have taken $\frac{3}{2}g_{\omega\Lambda}= g_{\omega N} = 12.5$ and $\frac{3}{2}g_{\sigma\Lambda}= g_{\sigma N} = m_N/f_\pi$.  The empirical values $m_N=939\,{\rm MeV}$, $m_\Lambda=1116\,{\rm MeV}$ and $m_\omega=783\,{\rm MeV}$ are taken from the particle data booklet. The scaling $\frac{g_\omega^*}{g_\omega}\approx 1-0.053 \frac{n-n_{/2}}{n_0}$ is taken as the ``best fit'' from the analysis in Section \ref{eos}. }\label{BS}
\end{table}
\begin{figure}[ht]
\begin{center}
\includegraphics[width=8.0cm]{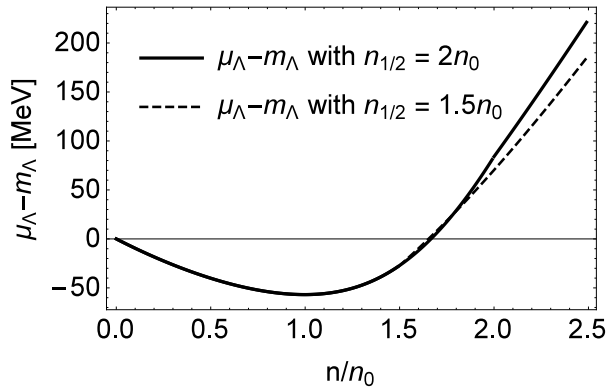}
\caption{$\mu_\Lambda -m_\Lambda$ vs. $n/n_0$ calculated with the scaling parameters determined in the theory and summarized in Table \ref{BS}. The demarcation density was chosen for $n_{1/2}= (1.5, 2.0)n_0$. The density at which the mass shift crosses zero corresponds to $\Delta$ of \cite{bedaque-steiner}.}
\label{mu_L_PKLR}
\end{center}
\end{figure}
Apart from the $\Lambda$ coupling to the mesons, the large cancelation between the $\sigma$ attraction and the $\omega$ repulsion responsible for small binding energy for nuclear matter must take place also in this case. In fact, using the standard constituent quark counting\footnote{In CT theory, the dilaton is a strong mixture of the quarkonium component and the gluonium component, so this quark counting may not be reliable. However we do not expect it to deviate much from 1.},  we may take $g_{\sigma \Lambda}\approx \frac 23 g_{\sigma N}$ and $g_{\omega \Lambda}\approx \frac 23 g_{\omega N}$, the 2/3 factor accounting for the two non-strange quarks in $\Lambda$ vs. 3 in nucleon. Then
\begin{eqnarray}
\mu_\Lambda
= m^*_\Lambda +\frac 23 \left(-\frac{g^{\ast\,2}_{\sigma N}}{m^{\ast\,2}_{\sigma}} n_s + \frac{g^{\ast\,2}_{\omega N}}{m^{\ast\,2}_{\omega}} n\right)\, \label{muL3}
\end{eqnarray}
This indicates that the $\Lambda$ effective mass shift $\mu_\Lambda-m_\Lambda$ will become positive near $\sim 2n_0$ as in the symmetric nuclear matter.

To make a rough estimate of the $\Lambda$ mass shift in dense medium, we take into account the d-scaling of the parameters in the $bs$HLS Lagrangian in the mean-field calculation which corresponds to  the ``single-decimation procedure" of \cite{BR:DD}. In doing this, it is important to recognize the scaling parameter  $c_I$ in this procedure could be different, i.e., renormalized, from the IDD coefficient entering into the double-decimation procedure with $V_{lowk}$ employed in Section \ref{eos}. The reason is that in the single-decimation procedure of RMF, as noted above, the scaling function $\Phi_I$ is related to the Fermi-liquid fixed-point parameters as shown in \cite{friman-rho} and encodes certain nonperturbative quasi-particle interactions on top of the IDD effects manifesting scale-chiral symmetry.
 We ignore this subtlety that we expect to be of higher order fluctuation corrections. Thus we take $c_{I}$ as used in the $V_{lowk}$ calculation, $c_{I}\approx 0.13$.  The scaling  and the constants for the dilaton and $\omega$ used in the calculation -- which are consistent with the property of nuclear matter treated in the mean field of the given Lagrangian -- are  summarized in Table \ref{BS}.

The result is plotted in Fig.~\ref{mu_L_PKLR}.  We see that $\mu_\Lambda-m_\Lambda$ crosses zero at a density $1.5 < n/n_0 < 2.0 $. The result is insensitive to the demarcation density for the regions. In fact what comes out in the mean field is quite easy to understand. Since $m^\ast_\Lambda$ stops dropping with $f^\ast_\sigma$ stabilizing at $2n_0$, what matters is the interplay of the ratio $(\frac{g^\ast}{m^\ast})^2$ for the scalar and vector mesons -- with an opposite sign -- multiplied, respectively, by the scalar density $n_s$ and by the baryon number density $n$. The vector repulsion wins over the scalar attraction as density increases in the same way as it does in nuclear matter. Although the estimate is admittedly approximate -- and it could be done much more realistically in the $V_{lowk}$ approach used above,  that the BS constraint~\cite{bedaque-steiner} is met is most likely robust. We conclude from this that within the formalism developed in \cite{LPR15} and with the prescription given in Ref. \cite{bedaque-steiner}, the hyperon problem does not arise in compact stars and hence the EoS discussed in Section \ref{eos} with hyperon degrees of freedom ignored could stay valid.
\subsection{Kaon condensation}
In the literature, the hyperon problem is treated independently of kaon condensation. We believe this is incomplete if not incorrect. In fact both hyperons and kaons figure together in flavor $SU(3)$ chiral Lagrangian and should be treated on the same footing. As will be discussed below, to ${\cal O} (N_c^0)$, hyperons and condensed kaons are likely to appear at the same density, with the possibility that higher-order corrections in $1/N_c$ could trigger  hyperons to appear {\it before} condensed kaons. In the preceding section, it was suggested that hyperons may be ignored in the EoS. The argument developed there was quite simple. In a close parallel to nuclear matter at high density where the repulsion between nucleons in exchange of $\omega$ mesons overpowers  the attraction due to scalar exchanges at densities near $2n_0$, $\Lambda$-nuclear interactions make the effective mass of a $\Lambda$ in medium  greater than that of a $\Lambda$ in the matter-free vacuum. In this section, we explore whether a related mechanism could be applied to avoid the ``kaon condensation problem."

\subsubsection{Callan-Klebanov skyrmion}
We first address the issue as to whether kaons condense before or after the appearance of  hyperons. At present, to the best of our knowledge,  the only way this problem can be addressed  in a tractable approximation in consistency with the basic premise of QCD -- such as large $N_c$ -- is the skyrmion description in which both kaons and hyperons can be  treated on the same footing with a same Lagrangian.  This matter was first discussed in \cite{LR-hk} employing the successful Callan-Klebanov bound-state model~\cite{callan-klebanov}. In this model, anti-kaons $K^{-}$ are bound to the $SU(2)$ skyrmion to yield hyperons. A highly non-trivial and surprising observation is that this model interpolates kaons between the chiral limit ($m_K\rightarrow 0$) and the Isgur-Wise heavy-quark limit ($m_K\rightarrow \infty$). The model can be applied to nuclear matter by putting the CK skyrmions on crystal lattice. It was shown~\cite{LR-hk} that put on a crystal,  the energy difference between the lowest-lying hyperon $\Lambda$ and the nucleon $N$ in medium comes out
 \be
E^\ast_\Lambda -E^\ast_N=\omega_K^\ast + {\cal O}(N_c^{-1})\label{inmedium}
\ee
where the asterisk represents medium-dependence. Note that the in-medium kaon mass is of $N_c^0$ in the $N_c$ counting. It is fortunate that the leading ${\cal O} (N_c)$ term and the flavor singlet ${\cal O} (N_c^0)$ Casimir energy term -- which is extremely difficult to calculate -- cancel out in the difference.

{Now, the $\Lambda$s will appear in the compact star matter when
\be
\mu_e\geq E_\Lambda^\ast -E_N^\ast
\ee
where $\mu_e$ is the electron chemical potential which is equal to $\mu_n-\mu_p$  in weak equilibrium}. On the other hand, kaons will appear by the weak process $e^-\rightarrow K^- +\nu_e$ when
\be
\mu_e\geq m_K^\ast.
\ee
Therefore to the leading order in $N_c$ in QCD, hyperons and condensed kaons populate compact stars simultaneously. Which one appears first in the single Lagrangian description depends on ${\cal O} (1/N_c)$ hyperfine corrections, namely, when the skyrmion-kaon system is rotationally quantized. A simple quasi-particle approximation leads to
\be
E_\Lambda^\ast-E_N^\ast=\omega^\ast_K +\frac{3}{8\Omega^\ast}({c^\ast}^2-1)\label{med-hyper}
\ee
where $\Omega>0$ is the moment of inertia of skyrmion rotator and $c^*$ is the in-medium hyperfine coefficient multiplying the effective spin operator of strangeness -1. The coefficient $c$ is highly model-dependent even in the matter-free space~\cite{callan-klebanov}, so it is  unknown in dense matter except in the large $N_c$ limit and also in the chiral limit. In either or both of these limits,  $c^\ast\rightarrow 1$. In the matter-free space, it is found to be $c^2\sim 0.5$. Although presently there is no proof, it seems likely that ${c^\ast}^2 <1$ in medium, approaching 1 from below near chiral restoration. If this is the case, that would suggest that hyperons appear before kaons condense\footnote{This conclusion is opposite to what was aimed at or conjectured by \cite{LR-hk}.} and they ultimately join in the vicinity of chiral restoration. It is however difficult to be more precise on this point since the effect is at ${\cal O} (N_c^{-1})$ and at that order there are many other corrections of the same order, such as higher-order nuclear correlations, that go beyond the mean-field order as in the $V_{lowk}$ approach of Section \ref{eos}. We are therefore unable to conclude that the absence of hyperons \`a la Bedaque and Steiner precludes kaon condensation.
\subsubsection{Relativistic Mean Field with $bs$HLS}
If hyperons do not figure in the EoS considered in Section \ref{eos} and kaons condense only after hyperons appear as suggested above, does it mean that kaon condensation can also be ignored? In order to address this question, we need a more detailed analysis within the framework developed in the paper. In the absence of $V_{lowk}^{SU(3)}$ approach, the best we can do is an RMF approach using the $bs$HLS Lagrangian with kaons implemented as ``heavy'' mesons with the scaling parameters in the $SU(2)$ sector fixed in Section \ref{eos}. A similar approach is discussed with dilaton treated differently from ours in \cite{chineseMF}.

In RMF, we can follow the argument given in \cite{BR-walecka-kcon}. In mean field, the kaon in medium receives mass shift by the tadpole in Fig.~\ref{tadpole} with the left baryon $(N,\Lambda)$ replaced by $K^-$. There is one striking difference between the baryon and the kaon. Unlike in the case of baryon where the scalar contribution is cancelled by the vector contribution, for the $K^-$,  both come in with the same sign thanks to the G-parity and the attractions add. This immediately precludes the mechanism that prevented the appearance of hyperons  for $n\gsim 2n_0$ for preventing kaon condensation.  This would mean that kaon condensation could intervene at higher densities where hyperons are not present.

What can prevent this was suggested in \cite{pandharipande-thorsson} where the authors
introduce higher-order nuclear correlations that involve repulsions in nucleon-nucleon interactions. The mechanism proposed in \cite{pandharipande-thorsson} is shown to push $n_K$ to $\gsim 6n_0$, above the central density of $\sim 2$-solar mass stars. This mechanism is not captured in the mean field with our $bs$HLS. It could however be captured in a three-flavor microscopic $V_{lowk}$ approach, a project relegated to a future research.
\vskip 0.3cm
$\bullet$ {\bf Other considerations}

(1) Of interest is the role of the IR fixed point of scale symmetry at which the dilaton mass is to vanish (in the chiral limit)~\cite{CT}. If the IR fixed point of scale symmetry is near $n_{cent}$, the possibility we have ignored in this paper, then one would have to consider the possible breakdown of Fermi liquid structure in the baryonic matter involved as discussed in \cite{PR-fermiliquid}. The break-down of Fermi liquid structure would make this problem a whole new ball-game. Kaon condensation in non-Fermi liquid is a totally unknown object.

(2) An alternative scenario is that if the kaon condensation threshold density is pushed up by the mechanism for the hyperon solution, then condensed kaons could be in a peaceful coexistence with non-Fermi liquid baryons in a form similar to strong-coupling strange quark matter co-existing with hadronic matter with no phase transition as discussed in a phenomenological model~ \cite{baym-kojo} or 3-layer structure consisting of hadrons, condensed kaons and strange quarks with a kaon-condensed state playing a doorway state to strange quark matter~\cite{KLR-3layer}.

\section{Conclusion}
\label{conc}
In this paper, we have constructed an  ``intrinsic density-dependent" scale-invariant hidden local symmetry (``$bs$HLS")  Lagrangian with baryons included explicitly, capturing sliding-vacuum properties induced by densities, and applied it to nuclear matter and compact-star matter.  In determining the ``bare" parameters of the $bs$HLS Lagrangian, we exploited the structure of baryonic matter present in the skyrmion description, namely, the skyrmion-half-skyrmion topology change at a density above that of the normal nuclear matter, and determined the IDD parameters in two regions of density R-I and R-II with the demarcation at the topology change density $n_{1/2}\approx 2n_0$.  It turns out, remarkably, that the scaling of the parameters of the ``bare'' Lagrangian with which the $V_{lowk}$ RG flow is to be performed can be put in the concise form $\frac{m_N^*}{m_N}\approx \frac{m_\sigma^*}{m_\sigma}\approx y_V \frac{m_V^*}{m_V}\approx \frac{\la\chi\ra^*}{\la\chi\ra}$ with $y_V=(\frac{g_V^*}{g_V})^{-1}$ and $V=(\rho, \omega)$. Apart from the quantity $y_V$ which is controlled by IDD$_{matter}$, the scaling of {\it all} light-quark hadrons in nuclear dynamics is dictated by IDD$_{pNG}$ representing the locking of chiral and scale symmetries.

With no unknown parameters, the properties of nuclear matter are well described by the $V_{lowk}$ RG approach up to the equilibrium density $n_0$ and are argued to be reliable in R-I up to the topology change density $n_{1/2}\approx 2n_0$.  In R-II, in contrast, due to the paucity of both theoretical and experimental input,  it is found to be difficult to pin down reliably the parameters of the bare Lagrangian. However relying on theoretical arguments based on the vector manifestation property of the $\rho$ vector meson and the pseudo-Nambu-Goldstone nature of $\pi$ and $\sigma$, we were able to fix almost all except for the $y_\omega$ for the $\omega$-NN coupling due to the apparent breakdown of $U(2)$ symmetry at high density. Assuming that the Fermi-liquid structure, known to be valid in the vicinity of nuclear matter density, continues to be applicable in R-II up the range of densities involved in compact stars, say, $\sim (5-6)n_0$, we were able to satisfactorily confront the properties of the recently observed massive neutron stars. We admit that were the Fermi liquid structure broken in the density regime concerned -- which cannot be excluded in the vicinity of the possible IR fixed point~\cite{PR-fermiliquid}, our results could not be trusted.

To summarize what we have found:

 The topology change that takes place in the skyrmion description of nucleons, incorporated into $bs$HLS Lagrangian, has a dramatic effect in the density regime $n>n_{1/2}\sim 2n_0$ on the EoS of compact-star matter. It affects the bare parameters of the effective Lagrangian due to the  existence of both the VM fixed point of the $\rho$ meson and the IR fixed point associated with the dilaton.
\begin{itemize}
\item It makes the nuclear tensor forces for $n\gsim n_{1/2} $ predominantly controlled by the pseudo-NG pion, with the competing $\rho$ tensor strongly suppressed, and induces a shift at $n_{1/2}$ from soft,  as needed in heavy-ion collisions~\cite{soft-heavy-ion}, to hard in the EoS, specially, the symmetry energy, as needed for massive neutron stars.
\item The changeover from skyrmion matter to half-skyrmion matter observed in the skyrmion-crystal description resembles, uncannily, the smooth transition at $n\sim (2-3)n_0$ from hadronic matter to strongly-coupled quark matter recently discussed~\cite{baym-kojo,hatsuda,fukushima-kojo}. In particular the half-skyrmion phase could be identified with the quarkyonic phase of \cite{fukushima-kojo}.
\item The $U(2)$ symmetry for $(\rho, \omega)$ which holds fairly well in the vacuum -- and presumably in R-I -- must be broken down at high density in R-II. Otherwise there will be inconsistency with the properties of the observed massive stars.
\item  The topology change with the consequent IDD parameter changes makes the $\omega$ repulsion dominate over the $\sigma$ attraction in R-II. This could potentially prevent the hyperons from appearing at a density $n\lsim 6n_0$, thus resolving the ``hyperon problem."  This could also be interpreted as the mechanism that accounts for the observed small binding energy -- an order of magnitude small relative to QCD scale -- for nuclei and nuclear matter, leading effectively to a BPS structure of baryonic matter discussed in \cite{LPR15},  seemingly at odds with QCD in the large $N_c$ limit. The same mechanism could prevent kaon condensations in the same range of density as that of hyperons but this requires further studies.
\item  What in our view is the most significant among our observations is the origin of proton mass as opposed to that of quark mass. The prediction of CT theory that the proton mass is dominantly gluonic, non-vanishing as the quark condensate goes to zero, and hence chirally invariant, is, albeit indirectly, supported by the results of this calculation. This suggests that the  mechanism for the proton mass generation lies outside of the standard paradigm based on spontaneous breaking of chiral symmetry. This feature is supported by skyrmion crystal models as well as parity-doubling baryon models.

\end{itemize}

Finally we should mention a fundamental issue related to the scale-chiral symmetry considered and its potential generalization.
In considering the scaling properties of baryons and mesons, we have implemented only the vector manifestation of $\rho$ which brings  (in the chiral limit) $\pi$ and $\rho$ into a zero mass multiplet. On the other hand, the scale-chiral symmetry considered in this paper implies the joining (in the chiral and scale limit) of $\pi$ and dilaton $\sigma$ into a zero-mass multiplet. The two could correspond to the same zero-mass multiplet. Together with the $a_1$, they could then constitute the multiplet figuring in Weinberg's mended symmetries~\cite{weinberg}. As discussed in \cite{LPR15}, a possible scenario could be that $\pi$, $\sigma$, $\rho$ and $a_1$ all come together in a massless multiplet at the chiral-scale restoration density. In the scaling behavior discussed above, $a_1$, not considered explicitly in this paper, could plausibly join $\pi$ and $\rho$~\cite{gen-hls} but there is no indication for the $\sigma$ dropping to zero within the range of densities involved. How and where they all tend to the mended symmetry limit, if such a limit exists, is not clear. This possibility contrasts with the supersymmetric QCD scenario (for $\rho$) of \cite{sqcd}.


\subsection*{Acknowledgments}
We thank  Yeunhwan Lim and R. Machleidt for many helpful discussions. One of the authors (MR) is grateful for instructive and stimulating comments from Rod Crewther, Lewis Tunstall and Koichi Yamawaki on scale invariance in both hadron and particle physics. The work of WGP is supported by the Rare Isotope Science Project of Institute for Basic Science funded by Ministry of Science, ICT and Future Planning and National Research Foundation of Korea (2013M7A1A1075764), that of TTSK in part by U.S. Department of Energy  under grant DF-FG02-88ER40388.
Part of this paper was written while two of the authors (HKL and MR) were visiting RAON/IBS for which the hospitality of Youngman Kim is acknowledged.

\newpage
\appendix
\centerline{\large\bf  APPENDIX}
\setcounter{section}{0}
\renewcommand{\thesection}{\Alph{section}}
\setcounter{equation}{0}
\renewcommand{\theequation}{\Alph{section}.\arabic{equation}}

\section{Density-Independence of the Pion Tensor Force }\label{appen_tensor}
In Section \ref{tensor_force} -- and in all previous works on tensor forces implemented with IDDs --  the density dependence of the pion tensor force was ignored, arguing that it is protected by chiral symmetry. Here we show explicitly that the argument is correct.

As shown in Section \ref{tensor_force}, the pion tensor  depends on density only via $m^\ast_\pi$.  To impose the scaling of $m^\ast_\pi$ consistently, we take in R-I
\begin{equation}
\frac{m^\ast_\pi}{m_\pi} = \Phi^{1/2}_I(n) \approx \left(\frac{1}{1+ 0.13 * n/n_0 } \right)^{1/2}\,.
\end{equation}
In R-II, the pion mass must drop fast since the bilinear quark condensate tends to zero, so it is reasonable to take it to decease rapidly and vanishing at the VM fixed point. Thus
\begin{equation}
m^{\ast\,2}_\pi \approx \frac{1}{f_{0\pi}^2\kappa^2} \sum_{n>1} c_n \langle \left( \bar{q}q \right)^n \rangle \sim \left( 1 - 0.15 * \frac{n}{n_0}\right)^2\, .  \label{mpiRII}
\end{equation}
The result is shown in Fig. \ref{pi_tensor}. We find that the pion tensor is more or less  independent of density both in R-I and in R-II although $m^\ast_\pi$ depends on density, where we used Fermi-Dirac distribution function as
\begin{eqnarray}
\frac{m^\ast_\pi}{m_\pi} &=& \left(\frac{1}{1+ 0.13 * n/n_0 } \right)^{1/2}\frac{1}{1+\exp\left(\frac{n-n_{1/2}}{0.05n_0} \right)} \nonumber\\
&&+ \left( 1 - 0.15 * \frac{n}{n_0}\right)\frac{1}{1+\exp\left(-\frac{n-n_{1/2}}{0.05n_0} \right)}
\end{eqnarray} to make $m_\pi^\ast$ be continuous at $n=n_{1/2}$. Thus taking the pion tensor density-independent in doing the $V_{low\, K}$ calculation in Section \ref{eos} is justified.
\begin{figure}[ht]
\begin{center}
\includegraphics[width=10.0cm]{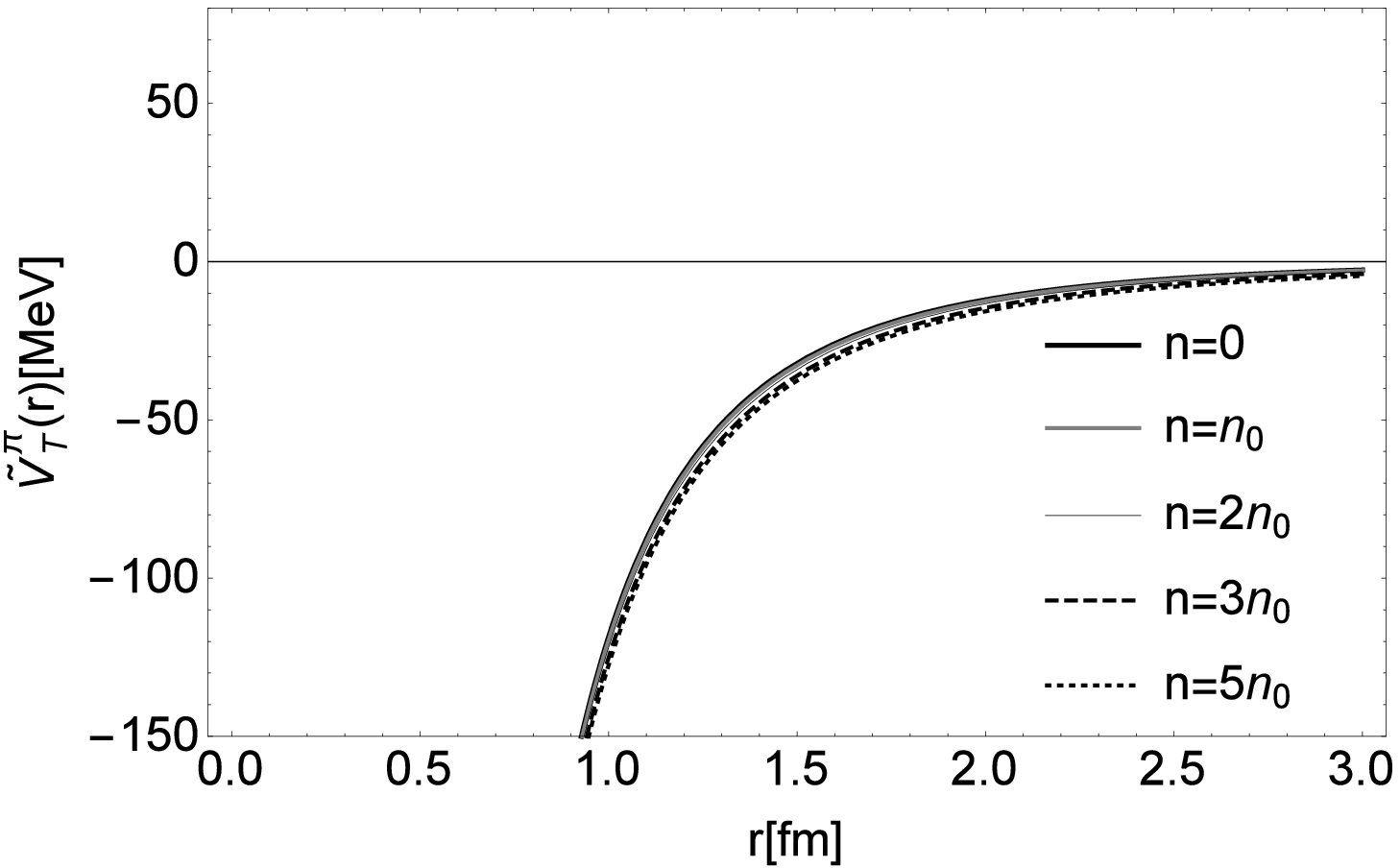}
\caption{ $\tilde{V}^\pi_T\left(r\right) \equiv V^\pi_T\left(r\right) \left( \tau_1\, \tau_2\,S_{12}\right)^{-1}$ with $n_{1/2} = 2n_0$. }
\label{pi_tensor}
\end{center}
\end{figure}

\section{One-Parameter Description of R-I}\label{AR-I}
In Section \ref{eos}, we have shown with a small fine-tuning for the constant $c_I$ consistent with the expected relation $c_I (N, \sigma)<c_I (\rho, \omega)$, nuclear matter can be reproduced well within the empirical error bars. Suppose one sticks to the one-$c_I$ strategy and asks how well nuclear matter can be reproduced. This has been checked for the range of $c_I =(0.13-0.20)$.  We see from Fig.~\ref{oneparameter} that taking a universal scaling parameter within a narrow range of the $c_I$ values fails to reproduce Nature. This may be due to different $1/N_c$ corrections contributing to the $c_I$ coefficients and  clearly indicates the extremely fine-tuned nature of the ground state property of nuclear matter.

\begin{figure}[h]
\begin{center}
\scalebox{0.70}{
\includegraphics[angle=0]{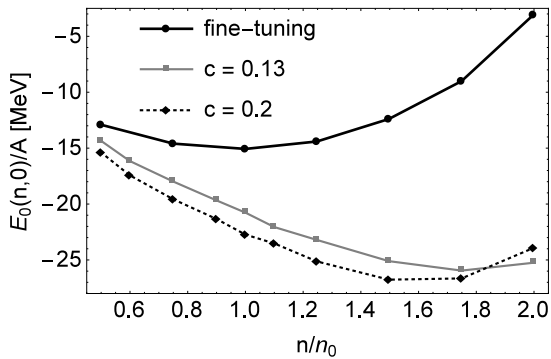}
}
\caption{Ground-state energy $E_0(n,0)$
of symmetric nuclear matter with one parameter $c_I$ in the range $0.13-0.2$.
 }\label{oneparameter}
 \end{center}
\end{figure}

\section{Fate of Hidden Local $U(2)$ Symmetry for $(\rho,\omega)$ in Region-II}\label{AR-II}
In Section \ref{region-II}, while local $U(2)$ symmetry was assumed to hold in Region-I (i.e., $n\leq n_{1/2}$), we suggested  that it should break down in R-II. There is no known theoretical argument either for or against this suggestion. Here we show an unequivocal indication from Nature that at least {\it within the present framework}   the symmetry should indeed break down in the density regime $n\gsim 2n_0$. The argument is based on the assumption that there is vector manifestation (VM) fixed point $n_{VM}\approx n_c \sim (6-7)n_0$ at which the $\rho$ mass vanishes (in the chiral limit).

Now let us suppose that the $U(2)$ symmetry holds in R-II. This would imply that near the chiral restoration point, the VM would hold for both $\rho$ and $\omega$.  We consider two possibilities: One scaling with the same slope in R-II and approaching the same point of VM
\begin{equation}
m^*_\omega/m_\omega\approx g^*_\omega/g_\omega\approx g^*/g \approx (1-n/n_c)\label{I}
\end{equation}
and the other approaching the VM fixed point with different slopes
{
\begin{equation}
m^*_\omega/m_\omega\approx \kappa\, g^*_\omega/g_\omega \approx \kappa \left(1-\frac{n -n_{1/2}}{n_c -n_{1/2}} \right).\label{II}
\end{equation}}
A drastic simplification is made on both and one should be cautious on the interpretation. Nonetheless,  the qualitative feature can be taken robust.
With all other parameters of Section \ref{eos} fixed the same, the ground-state energy of symmetric nuclear matter comes out as in Fig.~\ref{e0-nou2}.
\begin{figure}[h]
\begin{center}
\scalebox{0.48}{
\includegraphics[angle=0]{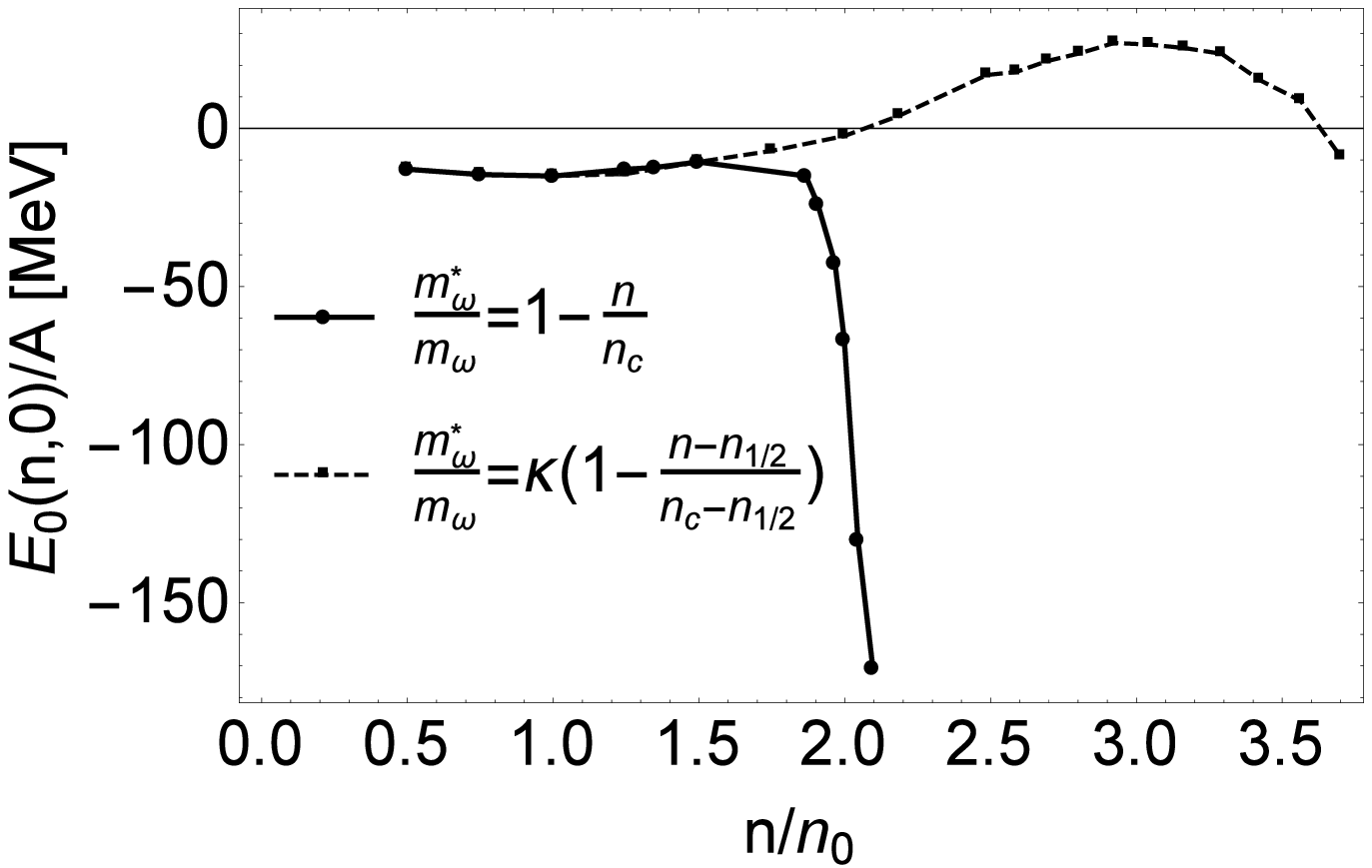}
\includegraphics[angle=0]{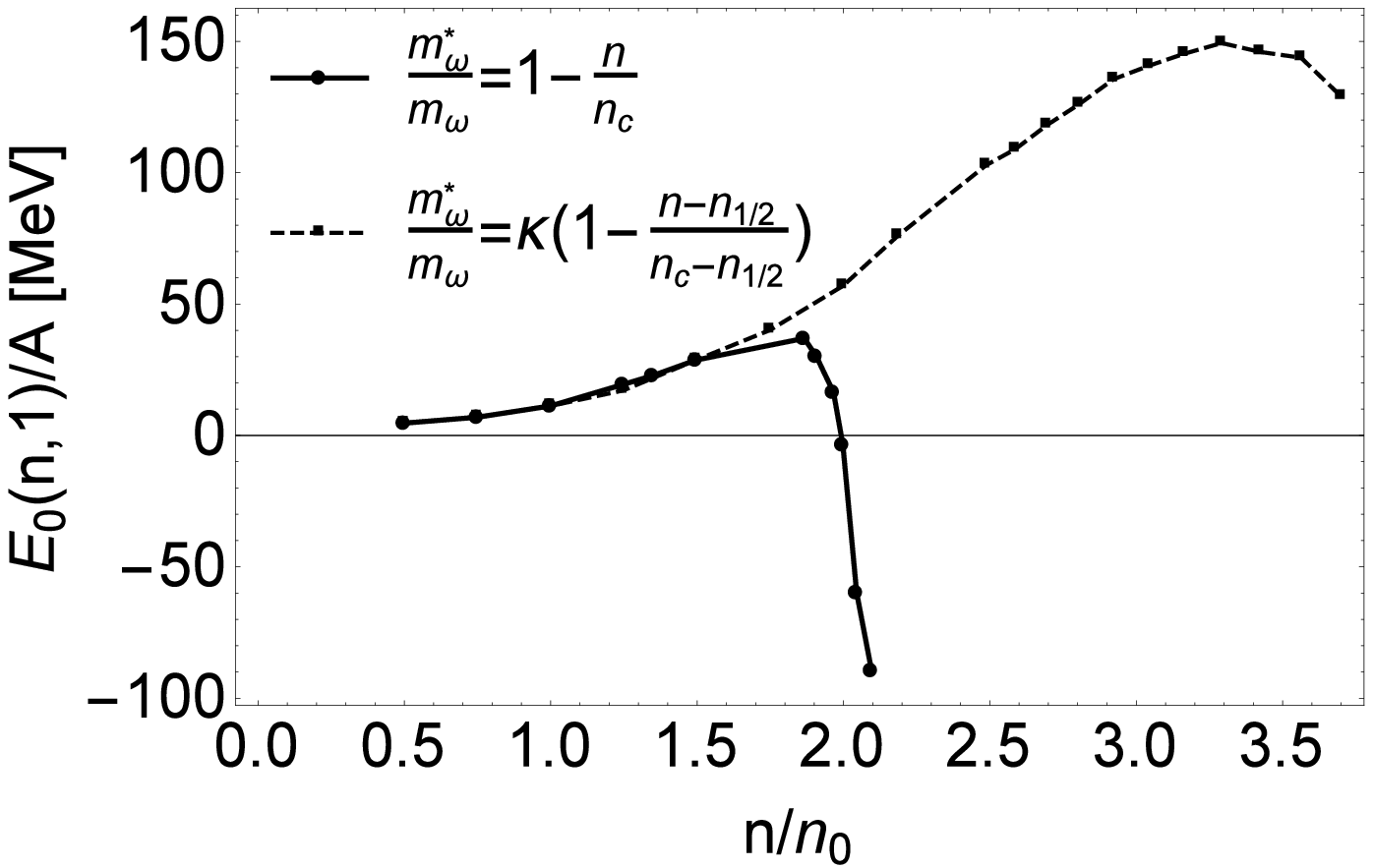}
}
\caption{Ground-state energy $E_0(n,0)$
of symmetric nuclear matter (left panel) and neutron matter (right panel) with $U(2)$ symmetry for $(\rho, \omega)$.
 }\label{e0-nou2}
 \end{center}
\end{figure}

One finds that both the symmetric matter and neutron matter become unstable at {$n\sim 2n_0$ for (\ref{I}) and $n\sim 3n_0$ for (\ref{II}).  This signals the breakdown. It takes place principally because the movement toward the VM fixed point softens the repulsion due to the $\omega$ exchange, the dropping $\omega$-nucleon coupling ``winning over" the increase in repulsion coming from the dropping mass. This makes the dilaton-exchange attraction taking over, leading to the collapse. Although the over-simplified linear d-scaling must bring about a precocious breakdown, this result indicates unequivocally that local $U(2)$ symmetry is untenable at high densities above $n_{1/2}$. We take this as a signal from compact stars that the hidden $U(2)$ gauge symmetry must be, perhaps badly,  broken at high density.


\end{document}